\documentclass[useAMS,usenatbib]{mn2e}
\usepackage{psfig}
\usepackage{graphicx}

\def\cbeta{$c_{\beta}$}  
\def\kms{\relax \ifmmode {\,\rm km\,s}^{-1}\else \,km\,s$^{-1}$\fi}

\def\mincir{\ \raise-2.truept\hbox{\rlap{\hbox{$\sim$}}\raise5.truept
    \hbox{$<$}\ }}
\def\magcir{\ \raise-2.truept\hbox{\rlap{\hbox{$\sim$}}\raise5.truept
    \hbox{$>$}\ }}

\def\arcmin{$'$}

\def\arcsec{\hbox{$^{\prime\prime}$}}
\def\nii{[N {\sc ii}]}
\def\hi{H {\sc i}}
\def\hii{H~{\sc ii}}
\def\sii{[S {\sc ii}]}
\def\siii{[S {\sc iii}]}

\def\oii{[O {\sc ii}]}

\def\heii{He{\sc ii}}
\def\hei{He{\sc i}}
\def\oiii{[O {\sc iii}]}

\def\ha{H$\alpha$}
\def\hb{H$\beta$}
\def\hd{H$\delta$}   
\def\hg{H$\gamma$}   

\def\te{$T_e$}
\def\tenii{$T_e$[N{\sc II}]}
\def\teoii{$T_e$[O{\sc II}]}
\def\tesiii{$T_e$[S{\sc III}]}
\def\teoiii{$T_e$[O{\sc III}]}

\def\ne{$N_e$}

\title[IC10 history: the chemistry of the PNe and HII regions]{IC10: the history of the nearest starburst galaxy 
through its Planetary Nebula and \hii\ region populations \thanks{Based on observations obtained at the Gemini 
Observatory, which is operated by the Association of Universities for Research in Astronomy, 
Inc., under a cooperative agreement with the NSF on behalf of the Gemini partnership.}}

\author[L. Magrini and D. R. Gon\c calves]{Laura Magrini$^{1}$\thanks{E-mail:
laura@arcetri.astro.it};
Denise R. Gon\c calves$^{2}$
\\
  $^{1}$ INAF - Osservatorio Astrofisico di Arcetri, Largo E. Fermi 5, I-50125 Firenze, Italy\\
  $^{2}$ UFRJ - Observat\'orio do Valongo, Ladeira Pedro Antonio 43, 20080-090 Rio de Janeiro, Brazil\\
}

\begin{document}

\date{Accepted ?. Received ?; in original form ?}

\pagerange{\pageref{firstpage}--\pageref{lastpage}} \pubyear{2008}

\maketitle

\label{firstpage}

\begin{abstract}
We report the results of spectroscopic observations, obtained with the
Gemini North Multi-Object Spectrograph, of 9 planetary nebulae (PNe)
and 15 \hii\ regions located in the 5.5\arcmin
$\times$5.5\arcmin\ inner region of the nearby starburst galaxy IC10.
Twelve new  candidate PNe have been discovered during our pre-imaging phase. Nine
of them have been spectroscopically confirmed.  The direct
availability of the electron temperature diagnostics in several
nebulae allowed an accurate determination of the metallicity map of
IC10 at two epochs: the present-time from \hii\ regions and the
old/intermediate-age from PNe.  We found a non-homogeneous
distribution of metals at both epochs, but similar average abundances
were found for the two populations. The derived age-metallicity
relation shows a little global enrichment interpreted as 
the loss of metals by SN winds and to differential gas outflows.
Finally, we analyzed the production of oxygen --through the third
dredge-up-- in the chemical abundance patterns of the PN populations
belonging to several dwarf irregular galaxies.  We found that the
third dredge-up of oxygen is a metallicity dependent phenomenon
occurring mainly for 12+$\log$(O/H)$\leq$7.7 and substantially absent
in IC10 PNe.
 \end{abstract}


\section[]{Introduction}

The irregular galaxy IC10 is the nearest starburst galaxy, sometimes
defined as the closest example of a blue compact dwarf 
\citep{richer01}.  It is located  in the outskirts of the Local Group
(LG), at low Galactic latitude.  Its location implies quite uncertain
determinations of its reddening and distance.  The estimates of the
reddening range from E(B-V)=0.47 to 2.0, while the distance modulus
from 22 to more than 27 mag, i.e. the linear distance of the galaxy 
would vary from 0.5 to 3~Mpc (Sakai, Madore  \& Freedman 1999 and Demers, 
Battinelli \& Letarte 2004, for a summary of distances and reddening).  
Recent estimations locate this galaxy at a distance around 0.7-0.8 Mpc 
(e.g. \citealt{demers04}; Kniazev, Pustilnik \& Zucker 2008; \citealt{sanna08}).

IC10 is remarkable for many reasons: {\em i)} its high star formation
rate (SFR), as evidenced by its large number of \hii\ regions
\citep{hodge90}, H$\alpha$ luminosity \citep{mateo98}, and far-IR
luminosity \citep{melisse94}; {\em ii)} its huge number of Wolf-Rayet
stars per unit luminosity, the largest in the LG (Massey, Armandroff 
\& Conti 1992; \citealt{massey95}), and the anomalous ratio of
carbon-type Wolf-Rayet stars (WC stars) to nitrogen-type Wolf-Rayet
stars (WN stars) which is peculiar at its metallicity (0.2--0.3 solar;
cf. Skillman, Kennicutt\& Hodge 1989 and \citealt{garnett90}); {\em
  iii)} the presence of an extended, complex and counter-rotating
envelope \citep{shostak89} revealed by \hi\ observations, which leads
\citet{wilcots98} to conclude that IC10 is still in its formative
stage.  All these characteristics suggest that IC10 is experiencing
an intense and very recent burst of star formation, starting about 10 Myr ago. 

Several studies in IC10 have revealed the existence of stellar
populations with varied ages from young to old  ages 
(\citealt{massey95}; \citealt{sakai99}; \citealt{borissova00};
Sanna et al. 2008, 2009), including also a number of PNe and carbon stars,
tracers of old/intermediate-age stellar populations
(\citealt{magrini03}; \citealt{demers04}).  However, with the
exception of the most recent star formation, very little is known about 
the star formation history of IC10 (but see the recent results by \citet{sanna09}) 
or its age-metallicity relation.

In this framework, the aim of the present paper is to study
spectroscopically two stellar populations of IC10: young stars
through the emission-line spectra of the \hii\ regions; and intermediate-age 
stars through PNe.  The
goal of this study is to reconstruct the star formation history of
IC10 from the birth of the PN progenitors to very recent times, and
to set up its age-metallicity relation.

This is possible thanks to the characteristics of the chemical
abundances derived from PN and \hii\ region spectroscopy.  The
progenitor stars of PNe, the low- and intermediate-mass stars (LIMS)
$1 M_{\odot}< M <8 M_{\odot}$, do not modify the composition of O, Ne
(these two elements in a first approximation), S, and Ar in the
material ejected during and after the Asymptotic Giant Branch (AGB)
phase.  Thus the PN abundances of these elements are characteristic of
the composition of the ISM at the epoch when the PN progenitor was formed.  
On the other hand, He, N, C are modified in the ejecta, because
processed during the lifetime of the progenitor stars.  They hence
give information on the stellar evolution of these stars. At the same
time, the chemical abundances of \hii\ regions are representative of
the current composition of the interstellar medium (ISM).  The
analysis performed in this paper will be the starting point of a
further work aiming at a complete understanding of the history of
IC10 through the building of an \lq ad hoc' chemical evolution model 
based on PN and \hii\ region constraints (Lanfranchi et al., in
prep.).

The paper is structured as follows: in $\S$ 2 we present the imaging
observations used for mask design, while in $\S$ 3 we describe our
spectroscopic observations with GMOS.  In $\S$ 4 we introduce the
determination of the physical and chemical properties of the nebulae,
discuss the nebular abundance patterns together with the age dating of the 
PN progenitors.  In $\S$ 5 we
analyze the star formation history of IC10, whereas in $\S$ 6 we
study the spatial metallicity distribution.  Finally, in $\S$ 7 we
approach the occurrence of the 3-rd dredge-up in PNe belonging to
several dwarf irregular galaxies.  In $\S$ 8 we present our summary
and conclusions.

\section[]{Pre-imaging}

GMOS-N pre-imaging exposures were taken in order to identify the
\hii\ regions and PNe selected to build the mask for our multi-object
spectroscopy, on August 08, 2007.  The central region of IC10, within
a field of view of 5.5\arcmin $\times$ 5.5\arcmin, was observed
through two filters: the H$\alpha$ one, HaG0310, with central
$\lambda$ 655 nm and width $\sim$7~nm and the H$\alpha$-continuum
filter, HaCG0311, whose central $\lambda$ is located in the continuum
adjacent to H$\alpha$ ($\lambda_c$ 662~nm, width $\sim$7~nm). The
exposure times were 400~s for each filter, split in two sub-exposures.
The two narrow-band frames were used to build a
\ha\ continuum-subtracted image, where we re-identified a large number
of small and giant known \hii\ regions, and three candidate PNe
discovered by \citet{magrini03}.  In addition, a large number of new
PN  candidates (12) were discovered.

The criteria adopted  to distinguish  PNe from compact HII regions are 
similar to those employed by Pe\~na, Stasi\'niska \& Richer (2007). In particular 
their criteria a), b), 
and d) are useful when only imaging observations are available, while 
e) is useful for spectroscopy. Their criterium c) is not applicable in our case.  
These criteria can be summarized in our case as the following: 
\begin{itemize}
\item[1)] at the distance of IC10 ($\sim$0.7-0.8 Mpc), 1\arcsec\ corresponds to about 3.6 pc.
Given the FWHM of about 0.6\arcsec\ for point-like objects, PNe (which have typically diameters 
smaller than 1 pc) are expected to be unresolved. Thus {\em our candidate PNe were selected among 
point-like \ha-emitting objects}; 
\item[2)] the central stars ionizing PNe are usually hotter than those ionizing \hii\ regions 
(e.g. \citealt{stas90}, \citealt{mendez92}). Due to their spectral type,  
with the energy maximum in the UV, PN central stars have lower $M_{\rm V}$  than the ionizing stars of \hii\ regions 
(typically more than 2 mag fainter). Thus {\em we do not expect to detect PN candidates in the image taken with  HaCG0311, 
whose central $\lambda$ is located in the continuum adjacent to H$\alpha$}.
\item[3)] Due to multiple ionizing sources in \hii\ regions, optically thick compact \hii\ regions 
are expected  to be brighter than optically thick PNe in the hydrogen recombination lines. Thus {\em our 
PN candidates are usually fainter in the \ha-continuum image than compact \hii\ regions}. 
\item[4)] From a spectroscopic point of view, the presence of  \heii\ 4686 emission whose intensity exceeds a few percent of H$\beta$
and/or intensity ratios \oiii~5007/H$\beta$ larger than about 4 are found only in PNe. On the other hand, low-excitation 
PNe cannot be distinguished from \hii\ regions only on the basis of their spectroscopic line ratios. 
\end{itemize}

Our further spectroscopic study confirmed 9 of these 12 objects as true PNe-- 
by the absence of continuum spectrum and/or from the high  \oiii~5007/H$\beta$ 
ratio (ranging from $\sim$4 to 16, with the exception of PN23, with a lower ratio). 
Among the point-like emission-line sources, we also revealed a symbiotic system
(IC10-SySt-01), confirmed spectroscopically. Its analysis has been
presented in a recent paper by \citet{goncalves08}. The remaining two candidate PNe 
were not included in the follow-up spectroscopy due to the intrinsic limits of the mask design.  

We used the standard star BD+254 to calibrate the \ha\ image. We compared the photometry 
of PN05, PN07, PN09 with that by \citet{magrini03}, finding an agreement with our measurements 
within $\sim$30\%.  The differences in the flux measurements are attributed to
the non-photometric conditions of the observations in the present work.

The identification numbers, equatorial coordinates at J2000.0 and \ha+\nii\ 
fluxes of the observed PNe and \hii\ regions
are shown in Table~1.  ID numbers in Table~1 were assigned starting
from the last number of the previous list of PN candidates in IC10
(ending with PN16; \citealt{magrini03}). Errors on the fluxes were
computed taking into account the uncertainties in the photon
statistics, the background, and the flux calibration.  The
PNe from 17 to 25 were spectroscopically confirmed, whereas the spectra of 
PN 26 and 27  were not observed.

\begin{table}
\centering
\begin{minipage}{80mm}
\caption{Identification and photometry of the \ha\ emitters. 
{\it Top}: known  PNe. {\it Middle}: new PNe candidates. {\it Bottom}: known \hii\ regions.}
\begin{tabular}{@{}llll@{}}
\hline
ID   &RA   &Dec      & H$\alpha$+\nii$^a$ \\
                    &J2000.0     &         &   \\
IC10PN5 &00:20:17.280&    +59:15:53.39& 1.68$\pm$0.20  \\  
IC10PN7 &00:20:22.232&    +59:20:02.57& 110$\pm$0.50   \\
IC10PN9 &00:20:32.113&    +59:16:02.45& 3.51$\pm$0.21   \\
\\
PN17 &00:19:57.487 &+59:17:12.77 &3.22$\pm$0.24\\ 
PN18 &00:19:59.418 &+59:19:03.53 &0.64$\pm$0.16\\ 
PN19 &00:20:03.103 &+59:19:20.53 &0.83$\pm$0.15\\ 
PN20 &00:20:03.896 &+59:19:27.41 &2.76$\pm$0.19\\ 
PN21 &00:20:14.798 &+59:18:08.96 &3.15$\pm$0.25\\ 
PN22 &00:20:19.712 &+59:18:13.81 &3.97$\pm$0.27\\ 
PN23 &00:20:23.695 &+59:19:51.99 &0.72$\pm$0.19\\ 
PN24 &00:20:29.507 & +59:16:41.38&3.72$\pm$0.29\\ 
PN25 &00:20:33.235 &+59:16:15.00 &1.19$\pm$0.15\\
PN26 &00:20:05.721 &+59:18:11.99 &1.14$\pm$0.15\\ 
PN27 &00:20:34.002 &+59:18:15.00 &2.23$\pm$0.16\\ 
IC10-SySt-01  &00:20:33.593 &+59:18:46.86 &1.36$\pm$0.16 \\
\\
HL90~13    & 00:20:04.270  &+59:16:56.53  & -             \\ 
HL90~20    & 00:20:10.025 & +59:19:14.92 &  -             \\ 	
HL90~29    & 00:20:12.864 & +59:20:09.05 & 102.3$^{b}$	  \\ 
HL90~30    & 00:20:13.299 & +59:20:13.36 &  109.6 $^{b}$  \\ 	    
HL90~40    & 00:20:16.719 & +59:20:28.92 &  9.55$^{b}$ 	  \\ 
HL90~45    & 00:20:17.439 & +59:18:40.14 &  1230.3$^{b}$  \\ 
HL90~50    & 00:20:19.256 & +59:18:55.30 &  794.3$^{b}$   \\ 
HL90~51    & 00:20:19.370 & +59:18:03.42    & 19.05$^{b}$ \\ 
HL90~107   & 00:20:25.921 &+59:16:49.71 &  57.5$^{b}$	  \\ 
HL90~111   & 00:20:27.584 & +59:17:25.643&  2818.4$^{b}$  \\ 
HL90~120   & 00:20:27.633 & +59:17:37.70 &  38.9$^{b}$    \\ 	     
HL90~122   & 00:20:27.887 & +59:18:21.28 &  42.6$^{b}$    \\ 	      
HL90~127   & 00:20:29.396 & +59:17:55.30 &  7.94$^{b}$    \\ 
HL90~141   & 00:20:32.462 & +59:17:50.337&  38.0$^{b}$	  \\ 
\hline
\multicolumn{4}{l}{$^a$ $\times$ 10$^{-16}$ erg cm$^{-2}$ s$^{-1}$.}\\
\multicolumn{4}{l}{$^b$ Fluxes from \citet{hodge90}.}\\
\end{tabular}
\end{minipage}
\label{tab_newPNe}
\end{table}

\section[]{Spectroscopy}

Spectra of IC10 PNe and \hii\ regions were obtained in queue mode
with GMOS-N, using two different gratings: R400+G5305
(`red'), with 3 exposures of 1,700s each, on October 11, 2007; and
B600+G5303 (`blue') with 4$\times$1,700s exposures, on October 14 and
18, 2007. The slit width was 1\arcsec,  while their heights varied from 
5\arcsec\ to 10\arcsec\ for PNe, and from 5\arcsec\ to 30\arcsec\ for \hii\ regions. 
Each exposure was offset by $\pm~3\arcsec$ from one another (positions A: -3$\arcsec$,B: 
0$\arcsec$, C: +3$\arcsec$, and D as A, for B600).
The pixel binning were 2$\times$2 (spectral$\times$spatial).  The spatial scale and
reciprocal dispersions of the spectra were as follows: 0\farcs094 and
0.3~nm per binned pixel, in `blue'; and 0\farcs134 and 0.8~nm per
binned pixel, in `red'. Seeing varied from $\sim$0.5\arcsec\ to
$\sim$0.6\arcsec\ for the R400 spectra, and it was
$\sim$0.6\arcsec\ for the two runs in which B600 spectra were
taken. CuAr lamp exposures were obtained with both gratings for
wavelength calibration.  The effective `blue' plus `red' spectral
coverage was generally from 3700~\AA\ to 9100~\AA. Due to the slit
location some spectra have a different spectral range (starting up to
300\AA\ above the lower limit, with the same $\Delta\lambda$, for
instance, from $\sim$4000 \AA\ to 9400~\AA).

Data were reduced and calibrated using the Gemini {\sc gmos data
  reduction script} and {\sc longslit} tasks, both being part of
IRAF\footnote{IRAF is distributed by the National Optical Astronomy
  Observatory, which is operated by the Association of Universities
  for Research in Astronomy (AURA) under cooperative agreement with
  the National Science Foundation.}.  
The sky- and background-subtraction task deserves a particular explanation, 
due to the difficulty of performing it in the central regions of IC10. 
The usual method consists in the  definition, for each slit,  of the object extension 
and one small (few arcsec) extension of the sky. In the central regions of the galaxy 
this was not possible since the slit was totally occupied by the emission-line source. 
Thus we allocated several slits in the external part of IC10, where no diffuse emission 
was present, to serve as sky template. We used them when the slit was completely 
filled by the source. 

The other method of sky- and background-subtraction takes advantage  
of three exposures in the `red' and four in the `blue' 
spectroscopic frames. In fact, the exposures  
were offset by $\pm~3\arcsec$ from one another (positions A: -3$\arcsec$,B: 0$\arcsec$, 
C: +3$\arcsec$, and D as A, but only for the blue part). As the sky transparency vary 
between the different exposures, we allowed a factor of normalization
adjusted in order to best remove the sky emission lines. 
The subtraction was performed taking into account the difference spectra: A-B, A-C, and B-C.
The final spectra were the average of the three individual pairs (A-B, A-C, and B-C) of sky-subtracted objects. 
Note that the extraction of all \hii\ regions has been limited to the bright central 2-3\arcsec region for obvious
reasons. The second method gives much better results in the sky-subtraction, especially in the near IR part of the
spectra, and therefore was the method used when possible, especially for PNe.
  
Spectra of the spectrophotometric
standard Wolf1346 (\citealt{massey88}; \citealt{massey90}), obtained
with the same instrumental setups as on September 17 and October 5,
2007, were used to calibrate the spectra.  This allowed to recover the
actual slope of the spectrum, although not its flux zero point. This
is not essential for our aims, for which only a relative calibration
in flux is necessary.

The emission-line fluxes were measured with the package {\sc splot} of
IRAF.  Errors on the fluxes were calculated taking into account the
statistical errors in the measurement of the fluxes, as well as
systematic errors of the flux calibrations, background determination,
and sky subtraction.

The observed line fluxes were corrected for the effect of the
interstellar extinction using the extinction law of \citet{mathis90}
with $R_V$=3.1.  We used \cbeta\ as a measurement of the extinction,
which is defined as the logarithmic difference between the observed
and theoretical \hb\ fluxes.  
Since \hd\ and \hg\ are only available
in few cases and are affected by larger uncertainties, \cbeta\ was
determined comparing the observed Balmer I(\ha)/I(\hb) ratio with its
theoretical value,  2.85 \citep{osterbrock06}.

Tables A.1 and A.2, in the Appendix section, give the results of the
emission-line flux measurements and extinction corrected intensities
for PNe and \hii\ regions, respectively.

The emission-line measurements of PN18 and PN19 are not presented in
Table~A.1. On one side, PN18 had its \hb\ emission line unmeasured because 
it fell within a CCD gap.  In addition, PN18 is a faint object and the only  
emission lines measurable in its spectrum are; \oiii\ 5007, \ha, and the \nii\ and \sii\ doublets.
On the other side, for the extremely
faint PN19 we could measure only the \oiii~5007\AA\ and \ha\,
lines. Instead of appearing in Table~A.1, the emission line
measurements of IC10PN7 --which turned to be an extended \hii\ region
composed by two knots, as already pointed out by \citet{kniazev08}--
are presented in Table~A.2.

We compared the observed emission line fluxes of Tables~A.1 and
Tables~A.2 with those emission line fluxes in common with previous
studies (\hii\ regions: \citealt{lequeux79} and \citealt{richer01};
PNe: \citealt{kniazev08}).  A good agreement is found not only for the
bright emission lines of \hii\ regions and PNe, but also for the
faintest cases, proving the reliability of the spectroscopy discussed here 
(see Figure 1).

\begin{figure} 
   \centering
   \includegraphics[width=8truecm]{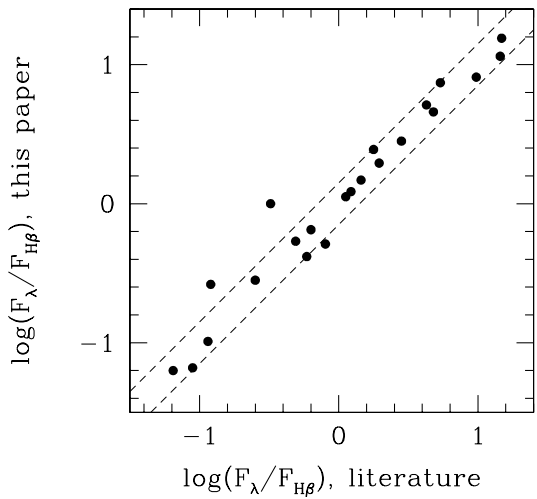} 
   \caption{Comparison of the observed flux measurements of the present work with previous 
   spectroscopic studies of the emission-line population of IC10 (\hii\ regions: \citealt{lequeux79} 
   and \citealt{richer01}; PNe: \citealt{kniazev08}).}
   \label{GMOS_im}
\end{figure}

\section[]{Physics and Chemistry of  the nebulae}

\subsection[]{Determination of nebular physical and chemical properties}

The extinction-corrected intensities were used to obtain the electron
densities and temperatures of each nebula, PN or \hii\ region.  To
calculate the densities we used the doublet of the sulfur lines
\sii$\lambda\lambda$6716,6731, while for the electron temperatures we
used the ratios \oiii $\lambda$4363/($\lambda$5007+$\lambda$4959)
and \oiii $\lambda$3727/($\lambda$7320+$\lambda$7330) and in few
cases \nii $\lambda$5755/($\lambda$6548+$\lambda$6584) and \siii
$\lambda$6312/($\lambda$9059 + $\lambda$9512).  The \nii\ and
\oii\ line ratio gives the low-excitation temperatures, while the
\oiii\ and \siii\ line ratios give the medium-excitation temperatures
(see also \citealt{osterbrock06}, $\S$5.2). Plasma diagnostics were
calculated using the 5-level atom model included in the {\sc nebular}
analysis package in {\sc iraf/stsdas} \citep{shaw94}.  The
\oiii\ $\lambda$4363 emission line was measurable with a sufficiently
high signal to noise ratio in 4 \hii\ regions and 1 PN, while an upper
limit to the \oiii\ ~$\lambda$4363 could be given for 6 \hii\ regions
and  5 PNe.  Due to the lower excitation of \hii\ regions respect to
PNe we could derive \teoii\ in 6 of them. 
We also could directly derive \tenii\ in HL90~45 and \tesiii\ in HL90~51.

The temperature and density uncertainties have been estimated by
formal error propagation of the absolute errors on the
extinction-corrected line fluxes.  Typical errors on electron
temperatures and densities are shown in Table~2 for nebulae with at 
least one electron temperature measured.  For the remaining objects,
electron temperatures are given in Table 3 and 4 as upper limits. 

Ionic abundances were computed using the {\sc nebular} analysis
package. Elemental abundances were then determined by applying the
ionization correction factors (ICFs) following the prescriptions by
\citet{KB94} for the case where only optical lines are available.

For the abundance analysis we used, when available, \teoii\ and
\tenii\ for the calculation of the N$^+$, O$^+$, S$^+$ abundances,
while \teoiii\ was used for the abundances of O$^{2+}$, S$^{2+}$,
Ar$^{2+}$, He${^+}$, and He$^{2+}$.  In the remaining objects, where
only \teoiii\ was measured, we adopted it both for low- and
high-ionization species.  The abundances of \hei\ and \heii\ were
computed using the equations of \citet{benjamin99} in two density
regimes, i.e. \ne\ $>$1000~cm$^{-3}$ and $\leq$1000~cm$^{-3}$.  The
Clegg's collisional populations were taken into account
\citep{clegg87}.

The remaining PNe and
\hii\ regions, those for which no estimations of electron temperature could be
given, are not reliable for any calculation of chemical
abundances.

In Table~2 we present the typical errors on the total chemical
abundances for nebulae with at least one direct estimate of 
electron temperature.  In this table \hii\ regions and PNe are grouped
in bins that correspond to their observed \ha\ flux in Table 1.  Errors on electron
temperatures and densities are given in percentage, while errors on He/H are
absolute and given on the quantity expressed by number. Finally, errors
on the total metal abundances are in dex on the quantities expressed
in the usual 12+$\log$(X/H) form.

In Tables~3 and 4 we present electron temperatures and densities, ionic abundances,
 ionization correction factors, and total abundances, for PNe and \hii\ regions, respectively. 
In these
tables, the quantities derived using an upper limit determination of
the electron temperature are marked  with  `$>$' for the forbidden lines and `$<$'  
for the recombination lines, according to their dependence on \te.

\begin{table*}
\centering
\begin{minipage}{180mm}
{\scriptsize  
\caption{Typical errors for nebulae with measured electron temperature. 
{\em a)} Errors on electron temperatures and densities are given  in percentage, 
{\em b)} errors on He/H are absolute and given on the quantity expressed by number, 
while {\em c)} errors  on the  total metal abundances are in dex on the quantities 
expressed in the usual 12+$\log$(X/H) form.}
\begin{tabular}{@{}llllllllllll@{}}
\hline
\ha\ flux     & n$_e^a$ & T$_e$[OII]$^a$ & T$_e$[OIII]$^a$ & $\Delta$(He/H)$^b$ & $\Delta$(O/H)$^c$& $\Delta$(N/H)$^c$& $\Delta$(Ne/H)$^c$& $\Delta$(Ar/H)$^c$& $\Delta$(S/H)$^c$ & N(PNe) & N(\hii) \\ 
10$^{-15}$ erg/cm$^{2}$ s    & \% & \% & \% & number & dex& dex& dex& dex&dex &  &  \\

\hline
1-10           &     -      & -                 & 12\%          &  0.03                 & 0.10                	& 0.15             	& -                       & -                   & 0.40  &1&-\\
10-50         &  25\%  & 9\% 	      & 12\%          &  0.008              & 0.08      			& 0.30   		& 0.40   			& 0.20   & 0.50  &-&2\\
50-100       &  20\%  &  -               &  10\%          &  0.007     		   & 0.07      			& -         		& -         			& -         & -        &-&1\\
100-1000   &  15\%  & 7\%           &  5\%            &  0.005  		   & 0.06     			& 0.15   		& 0.30   			 & 0.15   & 0.40  &-&4\\
$>$1000    &   10\% & 5\%           &   5\%           &  0.002  		   & 0.05      			& 0.15   		& 0.15  			 & 0.10   & 0.30  &-&2\\
\hline
\end{tabular}
}
\end{minipage}
\label{tab_err}
\end{table*}

\begin{table*}
\centering
\begin{minipage}{100mm}
{\scriptsize  
\caption{Physical and chemical parameters of the PN sample. $<$ and $>$ symbols mark the limits (upper or lower) on 
the total and ionic chemical abundance derived  with upper limit ($<$) determinations of the electron temperature.  }
\begin{tabular}{@{}lllllll@{}}
\hline
Id        &  IC10PN5  &         IC10PN9    &             PN17       &           PN20       &      PN21 &          PN22   \\    
\hline
Te(OIII)  &  $<$13000 &      $<$11000      &             14100      &         $<$10800     & $<$15000  &     $<$14700  \\    
Ne(SII)   &  300      &         8500       &      -                 &           3200       &  3100     &       4400    \\   
\hline
HeI/H     &  $<$0.137    &   $<$0.111            &       0.147            &     $<$0.086            &  $<$0.101   &    $<$0.060         \\   
HeII/H    &  $<$0.01    &     $<$0.003          &         0.009          &       -                         &  $<$0.007    &      $<$0.013       \\   
He/H      &  $<$0.149    &   $<$0.115            &       0.138            &     $<$0.086            &  $<$0.108    &    $<$0.072         \\   
OII/H     &  -           &   $>$5.67e-05        &       3.04e-05            &     -        	                   &  -                  &    $>$9.03e-05     \\   
OIII/H    &  $>$1.78e-04&    $>$3.29e-04       &        5.83e-05       &      $>$2.98e-04       &  $>$4.14e-05&     $>$1.37e-04    \\   
ICF(O)    &  1.05    &     1.02          &         1.02          &       1.00         &  1.01    &      1.02       \\   
O/H       &  $>$1.88e-04&  $>$3.93e-04         &      9.19e-05         &    $>$2.98e-04         &  $>$4.19e-05&   $>$2.33e-04      \\   
12+log(O/H)& $>$8.27  &   $>$8.59          &    7.96                &    $>$8.47                        & $>$7.62   &   $>$8.37        \\   
NII/H     &  $>$3.35e-06&    $>$5.36e-06       &      3.68e-06               &      -       	   &  -        &     $>$4.51e-06    \\   
ICF(N)    &  - 	      &     6.94         &         3.00          &       -              &  -        &      2.58       \\   
N/H       &  -        &    $>$3.72e-05       &        1.11e-05               &      -       	   &  -        &     $>$1.16e-05    \\   
12+log(N/H)& -        &   $>$7.57             &     7.05                 &   -   		   & -         & $>$7.06             \\   
ArIII/H   &  $>$5.46e-07&     $>$7.12e-07      &         -              &       $>$1.00e-06      &  $>$5.32e-07&      $>$6.59e-07    \\  
ICF(Ar)   &  1.87     &     1.17          &         -              &       1.87           &  1.87     &      1.635        \\  
Ar/H      &  $>$1.02e-06&  $>$1.33e-06         &      -                 &    $>$1.87e-06         &  $>$9.95e-07&   $>$1.23e-06       \\  
12+log(Ar/H)& 6.00:    &  $>$6.12              &    -                   &   $>$6.27               & $>$6.00      &  $>$6.09         \\  
SII/H     &  $>$8.69e-08&   $>$2.46e-07        &   7.74e-8                &     $>$6.33e-07        &  $>$5.17e-07&    $>$1.32e-07      \\  
SIII/H    &  $>$5.40e-07&    -               &        2.38e-06       &      $>$6.77e-07       &  $>$1.41e-06&     $>$8.58e-07     \\  
ICF(S)    &  -        &     1.39          &        1.05           &       -              &  -        &     1.09         \\  
S/H       &  -        &    $>$2.68e-06  	   &     2.49e-06          &          -           &  -        &  $>$1.08e-06        \\  
12+log(S/H)& -        &    $>$6.43            &     6.40               &       -              & -        &  $>$6.03           \\  
\hline
\end{tabular}
}
\end{minipage}
\label{tab_flux}
\end{table*}

\begin{table*}
\centering
\begin{minipage}{160mm}
{\scriptsize  
\caption{Physical and chemical parameters of the \hii\ region  sample. $<$ and $>$ symbols mark the limits (upper or lower) on 
the total and ionic chemical abundance derived  with upper limit ($<$) determinations of the electron temperature.   }
\begin{tabular}{@{}lllllllllll@{}}
\hline
Id            &  IC10PN7       & HL90~20     &  HL90~29         &    HL90~30   &HL90~45      & HL90~51     &  HL90~50   & HL90~107    &  HL90~111  & HL90~120             \\
\hline	                   		                    		                               					                 
Te(OIII)      &  $<$10600       &  $<$11000  &   14100          & $<$13500   	&10100         & $<$11600   &  12300     & $<$12350      & $<$10200  & 10300                   \\ 
Te(OII)       &  9500                & 8300             &  7500           &    -   		    &-             	    &-           	&  8600       & -                      &   9700         &  15000            \\      
Te(NII)       &  -                         &  -          	&   -              	&    -         		&11600            	    &-           	&  -               & -           		  &  -         & -                                                                     \\      
Te(SIII)      &  -                         &  -          	&   -              	&    -         		&-                  & -           	&  9200              & -                       &  -         & -                                                                                   \\      
Ne(SII)       &  180                   &  -          	&   -              	&    700       	&530            &600          	&  150         & 650                  &  -         &   613                                                         \\         
\hline	                   		                    		                               					                 
HeI/H         &  0.091         &  0.099      		&   0.132        	 &    $<$0.105     	&0.102         &$<$0.080       	&  0.121     & $<$0.091              &  0.086     & 0.054                                                                                         \\         
OI/H          &  3.24e-06     &  -          		&  -                	 &    -         		&1.25e-04   &$>$1.67e-06   	&  1.46e-06& $>$4.02e-06   &  -                  & 5.26e-07                                          \\  
OII/H         &  4.92e-05     &  9.44e-05  	&   3.994e-04     &    $>$1.06e-05  	&2.300e-05 &$>$2.24e-05   	&  8.77e-05& $>$1.54e-05   &  6.218e-05 & 4.381e-06                                                         \\  
OIII/H        &  1.16e-04     &  8.702e-05  	&   8.89e-05       &    $>$4.44e-05  	&2.613e-04 &$>$8.94e-05   	&  6.6e-05 & $>$1.40e-04   &  6.249e-05 & 1.38e-04                                                 \\                                                      \\ 
O/H           &  1.69e-04     &  1.82e-04  	&   4.88e-04       &    $>$5.53e-05  	&2.87e-04   &$>$1.14e-04   	&  1.56e-04 & $>$1.61e-04   &  1.264e-04 & 1.53e-04                                                   \\       
12+log(O/H) &  8.23         &  8.26      	&   8.68           	 &   $>$7.74     	 &8.45           &$>$8.05       	&  8.19      & $>$8.21        &  8.10               & 8.18          \\    
NII/H         &  5.44e-06     &  5.675e-06  	&   2.00e-05       &   $>$2.44e-05   	&1.315e-06  &$>$3.0e-05   	&  4.950e-06 & $>$2.39e-06   &  5.858e-06 & 1.44e-06                                                                \\      
ICF(N)        &  3.43        &  1.92     		 &   1.22              &    5.2        	&12.5             &1.55       	&  5.08       & 10.5       &  2.032                 & 34.9                                                            \\        
N/H           &  1.87e-05     &  1.09e-05  	&   2.45e-05       &    $>$1.2e-05    	&1.64e-05     &$>$1.53e-04   &   8.8e-06 & $>$2.51e-05   &  1.191e-05    & 5.04e-05                                                 \\         
12+log(N/H) &  7.27         &  7.03      	&	  7.40         &    $>$8.10     	&7.21              &$>$8.18      	&  6.95      & $>$7.40        &  7.08               & 7.70        \\     
NeIII/H       &  1.01e-05     &  -          		&   -              	&    -           	&4.69e-05     &-           	&  9.26e-06 & -           &  -                         & 1.52e-05               \\      
ICF(Ne)       &  1.46         &  -         		 &   -              	&    -            	&1.09             &-           	&  2.36     & -           &  -                             & 1.10                                        \\ 
Ne/H          &  1.47e-05     &  -          		&   -              	&    -            	&5.14e-05     &-           	&  2.18e-05 & -           &  -                        & 1.68e-05                         \\       
12+log(Ne/H)&  7.17        &  -          		&   -              	&    -            	&7.71       	      &-           	&  7.34     & -           &  -                             & 7.22           \\
ArIII/H       &  9.23e-07     &  0.87e-06  	&   0.86e-06       &    $>$0.57e-06    &1.13e-06     &$>$8.00e-07   &  7.45e-06 &$>$ 0.950e-06   &  1.100e-06 & 0.62e-06      \\  
ICF(Ar)       &  1.41         &  2.1      		&   5.5               	&    1.23        	&1.08             &1.24       	&  2.28     & 1.10       &  1.969                   & 1.03                                       \\
Ar/H          &  1.73e-06     &  1.58e-06  	&   1.60e-06      &       $>$1.06e-06   &2.13e-06     &$>$1.49e-06   &  1.39e-06 & $>$1.78e-06   &  2.057e-06 & 1.23e-06    \\ 
12+log(Ar/H)&  6.24         &  6.20          	&   6.20       	&    $>$6.03       	& 6.32             &$>$6.17       	&  6.14      & $>$6.25        &  6.31                  & 6.10          \\
SII/H         &  4.34e-07     &  6.80e-07  	&   3.65e-06      &     $>$3.12e-07     &2.31e-07     & $>$4.26e-07  &  6.221e-07 & $>$0.69e-06   &  7.446e-07 & 1.67e-07    \\
SIII/H        &  3.690e-07     &   2.13e-06 	&   1.90e-06      &    $>$2.090e-06    &1.74e-06     & $>$2.76e-06  &  3.46e-06 & $>$5.06e-06   &  -                    &  1.52e-06                                                    \\
ICF(S)        &  1.26         &   1.04     		&   1.010          	&    1.28          	 &1.65              & 1.27      	&  1.029     & 1.57       &  1.048                  & 2.28                                    \\
S/H           &  4.08e-06     &  2.92e-06  	&   2.26e-06      &    $>$3.02e-06       &3.27e-06     & $>$4.07e-06  &  4.20e-06 & $>$9.03e-06   &  5.212e-06 & 3.88e-06     \\
12+log(S/H) &  6.67         &  6.46      	& 7.35              	&    $>$6.48     	& 6.51              &  $>$6.61   	&  6.62      & $>$6.95        &  6.72                 & 6.59       \\    
\hline
\end{tabular}
}
\end{minipage}
\label{tab_flux}
\end{table*}

\subsection[]{Abundance patterns: PNe and \hii\ regions}
\begin{figure}
\centering
\includegraphics[width=8.5truecm]{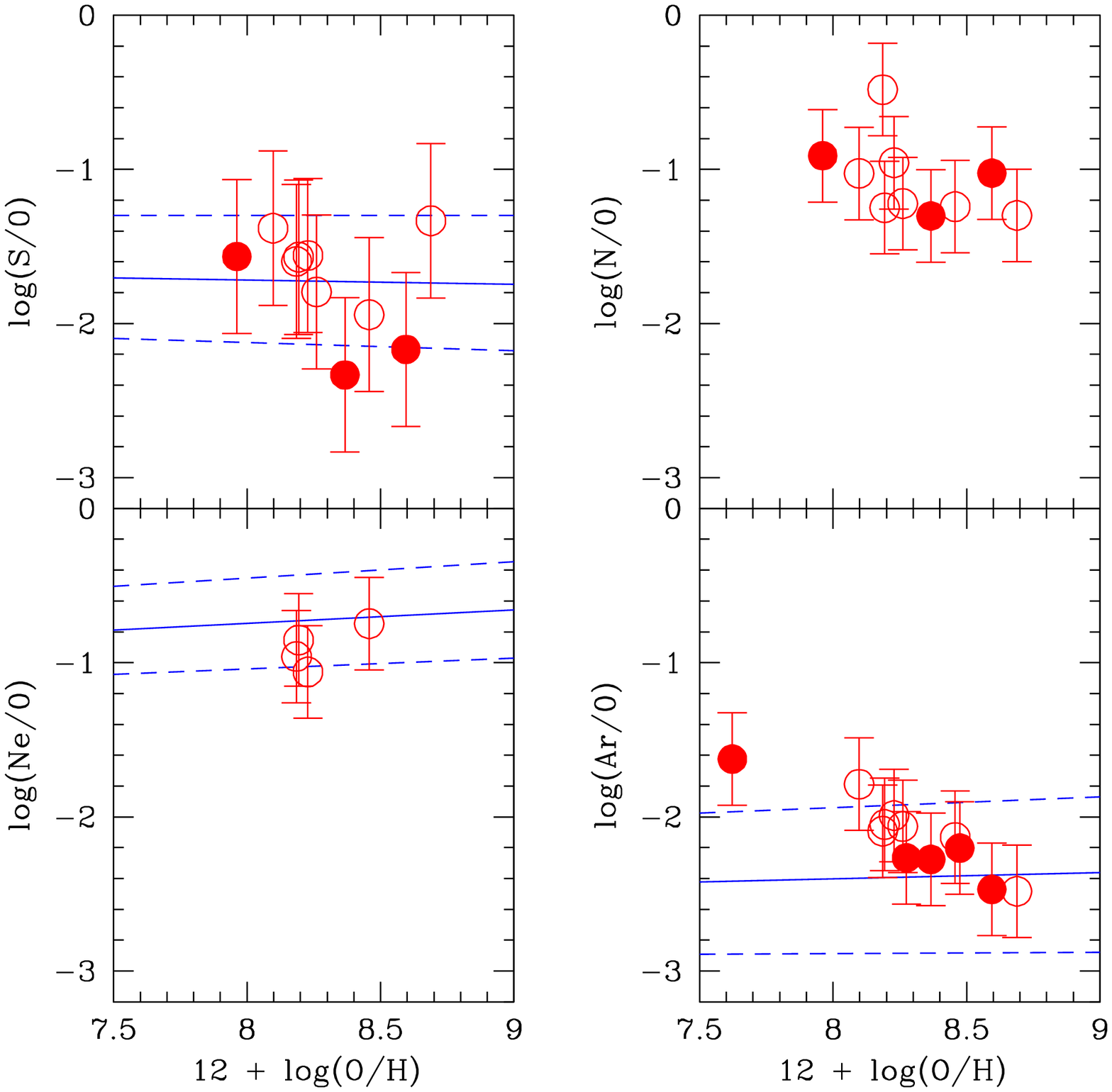} 
\caption{Chemical abundance patterns for PNe (filled circles) and \hii\ regions (empty circles), 
compared with the relations of \citet{Izotov06} for the $\alpha$-element to oxygen ratios.  
The Izotov's relations are marked with continuous lines, while their dispersion is indicated 
with dashed lines.}
\label{pat}
\end{figure}

We compared the abundance ratio of PNe and \hii\ regions of IC10 with
the relations derived by \citet{Izotov06} from observations of a
large sample of blue-compact galaxies, BCGs. These relations are marked by 
continuous
lines in Figure~2, while their 1-$\sigma$ dispersion is indicated by
two dashed lines.  In the Izotov et al.'s sample, the $\alpha$
element-to-oxygen abundance ratios Ne/O, S/O and Ar/O do not show
large trends with oxygen abundance, due to their common origin.  The
best determined ratio, Ne/O, increases slightly with increasing O/H
(by $\sim$0.1 dex in the analyzed range).  This can be explained by
the depletion of oxygen onto dust grains with $\sim$20\% of oxygen locked
in dust.  Due to the large dispersion of N/O and the different stellar
origin of nitrogen and oxygen, \citet{Izotov06} do not present a
relation of this ratio with 12+log(O/H) (see, e. g., Figure~11 of
Izotov et al. 2006 and the large scatter for 12+$\log$(O/H)$>$8).

The abundance ratios of IC10 \hii\ regions are generally in good
agreement with the trends of BCGs.  The Ne/O ratio is
measured in 4 \hii\ regions and shows little dispersion: 
Ne/O=0.13$\pm$0.04, which means $\log$(Ne/O)=-0.89$\pm$0.10dex.  It is
slightly lower than the value found by \citet{Izotov06}, but still
consistent with their relation.

The values of $\log$(N/O) in the \hii\ region sample show a larger
dispersion (-1.0 $\pm$ 0.3) than Ne/O, but they are consistent with
the values assumed by BCGs in the same metallicity range (see Figure 11
of Izotov et al. 2006, where for 12+$\log$(O/H)=8.3, the
corresponding $\log$(N/O) ranges from approximately -1.3 to -0.9).
However, considering the recent star formation history of IC10, which is
having a strong starburst, we would expect to measure a lower value of
N/O.  In fact, in the delayed-release hypothesis, i.e. the
moderate-time-delayed production of nitrogen by intermediate mass
stars \citep{pi03}, N/O would drop while O/H increases as massive
stars begin to die and eject oxygen into the interstellar medium.  At
the point when all the massive stars have died, the N/O value would be
at its minimum.
The reason why the N/O value in IC10 is consistent with other BCGs has to be 
searched in the fact that anything ejected by massive stars  
--whether oxygen by type II supernovae or nitrogen in the winds of WR stars--, 
will not be immediately incorporated into the nebular gas.  In fact, this ejected matter is
too hot, and needs to cool before it can mix with the ISM and be seen
in the optical spectra of nebular gas.
In addition, we note that there is a single
\hii\ region with a higher N/O value, namely HL90-120.  We thus
consider for this region, which contains four known W-R stars within
20\arcsec\ from its center, a possible zone of nitrogen pollution. A
large presence of Wolf-Rayet stars could, indeed, produce isolated
regions with enhanced nitrogen \citep{lopez08}.

For the  three PNe for which we measured the N/O ratio, we obtain values
close to those expected for the \hii\ regions in BCGs.  
From the nucleosynthesis of the PN progenitors, one would expect 
an enhancement of the N/O ratio since PNe commonly synthesize and
dredge-up nitrogen.  
Recently, \citet{RM07} have shown that, often, 
the brightest PNe in star-forming galaxies have N/O ratios very similar to those in the
\hii\ regions.
The values of both Ar/O and S/O of \hii\ regions
and PNe are in agreement with the Izotov et al.'s (2006) 
relations. The decreasing trend of Ar/O vs. O/H might be due to the
adopted ICF for Ar.
\subsection[]{The nature of the PN population: dating the progenitors}

The abundance ratios N/O and He/H are useful to discriminate PNe of
different types, which means PNe deriving from progenitors with
different initial mass and thus formed in different epochs. The N/O
ratio provides information about the stellar nucleosynthesis during
the AGB phase of LIMS. In fact, nitrogen is produced in AGB stars in
two ways: by neutron capture, during the CNO cycle; and by hot-bottom
burning. Hot-bottom burning produces primarily nitrogen but occurs
only if the base of the convective envelope of the AGB stars is hot
enough to favor the conversion of $^{12}$C into $^{14}$N.  Thus
nitrogen is expected to be mostly enriched in those PNe with the most
massive progenitors, i.e. with turnoff mass larger than $\sim$3
M$_{\odot}$ (\citealt{van1997}; \citealt{marigo01}).  The He/H abundance 
gives also an indication of the initial mass of the progenitor: the
nebula is enriched progressively for more massive stars; it reaches a
plateau between 3 and 4 M${_\odot}$; and then it increases again
toward the higher masses \citep{marigo01}.  However, He/H is a weaker
indicator of the PN types in low metallicity environments.

Since the oxygen abundance of IC10 is very similar to those of the LMC
and of the M33, we can use the Type I limits as in \citet{dopita91} and
\citet{magrini09}: Type I PNe are those with
log(N/O)$>$-0.5 independently of the helium abundance.  From Table~3 we
see that N/H is available for three  PNe, namely IC10-PN9, PN17, and 
PN22. Following the definition by \citet{dopita91} these PNe are
non-Type I. For the remaining PN we have not a sure determination of
their type.
Since the  N/O ratio is low for all three PNe, high mass progenitor 
stars can be ruled out for them. Therefore, their progenitor stars are  
of mass,  M $<$1.2 M$_{\odot}$, and consequently they  were born during 
the first half of the age of the Universe.

For the PNe without any determination of the N/O ratio, another aspect 
of their spectra points out that they have old progenitors. The \heii\  
intensities are all very low or not detectable, which implies that the 
central stars are on the horizontal portion of their evolutionary tracks.  
This can happen irrespective of the age/mass of the progenitor stars, but it is more likely if 
the stars are of low mass since the PNe will have lower mass central stars that evolve to 
high temperature more slowly.

\subsection{The chemical evolution of IC10 and comparison with other galaxies}

\begin{table*}
\centering
\begin{minipage}{150mm}
{\scriptsize  
\caption{Average chemical abundances and abundance ratios of PN and \hii\ region samples in several galaxies. 
The first column indicates the sample; columns from 2 to 7 show the average elemental abundances expressed by 
number; columns 8 and 9 give the mean values of N/O and Ne/O. {\em a)} Chemical abundances computed 
also including upper limit electron temperature determinations; {\em b)} chemical abundances computed 
using only direct electron temperature determinations; {\em c)} \citet{stang08}; 
{\em d)}; \citet{magrini09} {\em e)}~\citet{stang06}; {\em f)} Asplund, Grevesse \& Sauval
(2005).}
\begin{tabular}{@{}lllllllllll@{}}
\hline
Sample             & He/H  & O/H  & N/H  & Ne/H  & Ar/H   & S/H  & N/O & Ne/O \\
                   &       & ($\times10^{-4}$) & ($\times10^{-4}$) & ($\times10^{-5}$) &  ($\times10^{-6}$)  &  ($\times10^{-6}$) & &  \\
\hline
IC10 PNe$^a$             &0.107$\pm$0.032   & 2.07$\pm$1.30 &0.20$\pm$0.15   &-                           & 1.29$\pm$0.35 &2.08$\pm$0.87 &0.09$\pm$0.04 &  -\\
IC10 PN-17$^b$         &0.147$\pm$0.03   & 0.91$\pm$0.1 &0.11$\pm$0.02  &-                           & - &2.5$\pm$0.8 &0.11$\pm$0.1 &  -\\
\\ 
SMC PNe$^c$          & 0.113$\pm$0.022 &1.05$\pm$0.46  & 0.28$\pm$0.33 &  1.77$\pm$1.32& 0.59$\pm$0.59 & 4.80$\pm$6.57 & 0.28$\pm$0.50 & 0.17 $\pm$0.08\\
LMC PNe$^c$          & 0.103$\pm$0.026 &2.32$\pm$1.65  & 1.48$\pm$1.75 &  4.04$\pm$3.60& 1.14$\pm$0.72 & 3.46$\pm$8.88 & 0.87$\pm$1.15 & 0.17 $\pm$0.09\\
M33 PNe$^d$           & 0.118$\pm$0.075 & 2.33$\pm$1.58 & 1.40$\pm$2.49 & 4.91$\pm$4.14 & 1.20$\pm$0.57 & 5.91$\pm$3.58 & 0.40$\pm$0.31 & 0.17$\pm$0.06 \\ 
Galactic PNe$^e$    & 0.123$\pm$0.042 & 3.53$\pm$1.95 &  2.44$\pm$3.46 & 9.68$\pm$7.98 & 1.26$\pm$1.24 &     -                       &   0.67$\pm$0.82 &  0.25$\pm$0.10        \\
\\ 
IC10 HII regions$^a$  &0.107$\pm$0.031      & 2.31$\pm$1.59      &0.38$\pm$0.39     &2.62$\pm$1.76       & 2.49$\pm$2.81       &7.45$\pm$6.62                       & 0.10$\pm$0.10         & 0.13$\pm$0.04\\
IC10 HII regions$^b$   &0.100$\pm$0.024     & 2.23$\pm$1.28      &0.20$\pm$0.14     &2.62$\pm$1.71       & 1.67$\pm$2.81       &6.68$\pm$7.10       & 0.11$\pm$0.10    & 0.13$\pm$0.04\\
\\ 
M33 HII regions$^d$    & 0.101$\pm$0.015 & 2.04$\pm$0.75 & 1.28$\pm$0.57 & 4.24$\pm$2.61 & 1.31 $\pm$0.45 & 5.85$\pm$2.28 & 0.06$\pm$0.02 &0.20$\pm$0.06   \\
\\ 
Solar value$^f$      & 0.085$\pm$0.02       & 4.57$\pm$0.04       & 0.60$\pm$0.09      & 6.9$\pm$1.0   & 1.51$\pm$0.30 & 13.8$\pm$2.0  & 0.13$\pm$0.10 & 0.15$\pm$0.05\\
\hline
\end{tabular}
}
\end{minipage}
\label{tab_avabu}
\end{table*}

In Table 5 we compare the average chemical abundance of the PN and 
\hii\ region populations of IC10 with those of SMC, LMC, M33, and the Milky Way 
galaxies. The uncertainties given in Table 5 are the standard deviations 
of the average values, computed by number and then converted in the 
logarithmic form. The large {\em rms} uncertainties both in PN and \hii\ 
region average abundances are  due to their spatial variations through  
the disk of IC10.  

First of all, by comparing the average abundances of the $\alpha$-elements of the
PNe of IC10 with those of its \hii\ regions, we found that they nearly 
match, with the PN abundances slightly lower. 
In particular,
S/H and Ar/H are lower in PNe than in \hii\ regions, while O/H is
unchanged, within the errors. Note that oxygen is the best measured
element and at the metallicity of IC10 it is not modified during the
lifetime of LIMS (see the detailed discussion in $\S$7). Thus, we can
adopt it as a tracer of the past ISM metallicity.  
At a first look the enrichment history of IC10 can be thus estimated by 
comparing the average O/H of \hii\ regions, e.g. the present-time ISM, 
to the O/H of PNe, the ISM at the epoch of the birth of the progenitor 
LIMS.  From this comparison, we found a small variation in the average
metal content of IC10, from the epoch of the formation of the PN
progenitors (see Sect. 4.3)  to the present time.  However, we remind that only a
single PN has a direct measurement of the electron temperature.  The
average PN abundance in Table~5, marked with {\em a}, also includes
upper limit electron temperatures, i.e., lower limit oxygen
abundances.
Unfortunately, no abundance determination from other stellar
populations than \hii\ regions are available in the literature.
Previous chemical abundances were based on spectroscopic observations
of \hii\ regions by \citet{lequeux79}, then recomputed with updated
atomic data and ICF by \citet{skillman89} and \citet{garnett90}, and
more recently by new spectroscopic observations of Richer et
al. (2001). These abundance determinations are in good agreement with
ours, but do not extend the temporal range.
The reasons for the little change in the global metallicity of IC10 have to be searched 
in the strong winds and outflows that are affecting its chemical evolution. 
These winds are believed to be strongly differential and to allow mainly the lost of 
$\alpha$-elements expelled  through Type II supernova winds (e.g, \citealt{maclowferrara99}, 
\citealt{recchi08}).  

Secondly, let's compare the chemical abundances of PNe in different galaxies: 
the metallicity of the PNe in IC10 is sub-solar, and their average 
oxygen abundance is very close to those of the LMC and M33. 
While a good agreement is noticed among PNe $\alpha$-element abundances,  
both N/H and consequently N/O are lower than the corresponding 
values in LMC and M33.
 
Finally, by examining  the \hii\ regions we find again a good agreement 
with M33, with N/O ratio consistent 
with the average value of M33's \hii\ regions.  
Thus, it seems that the main difference between the average abundances (PN and \hii\ regions)   
of IC10 and M33 is in the behaviour of the N/O ratio: 
while N/O is similar in the \hii\ regions of both galaxies, it results to be higher in  M33 PNe 
than in IC10 PNe. Is this a signature of different past star formation histories or a lack of nitrogen 
enrichment in the PN progenitors? Probably the presence of a number of Type I PNe in M33, which 
are indeed absent in IC10, enhance the N/O average value. As in other dIrrs, many of the 
progenitors of PNe do not show evidence of nitrogen dredge-up, which might be a uncommon 
process in such low metallicity environments.
An interesting discussion on the N/O and the nucleosynthesis processes in the
progenitors of PNe in dIrr galaxies is treated by \citet{RM07}. Considering a sample 
of bright PNe (i.e., within 2 mag from the bright cutoff of their luminosity function)
in several nearby dIrrs they found that PNe have oxygen abundances close to the ISM 
values and also N/O has little variations in the two populations. Since the Ne/O ratio 
is in agreement with the ISM one, they conclude that nor oxygen neither neon are modified 
during the PNe lifetime. Thus, since O/H is close the \hii\ regions' one, they conclude 
that these PNe  arise preferentially as a result of a recent star formation. But in IC10, 
this conclusion is difficult to reconcile with the estimated large age of PNe, both from 
their low \heii\ lines and their low N/O ratio, which are both hints of low mass (thus old) 
progenitors. In IC10, the characteristics of PNe are more consistent with an 
old/intermediate-age population formed from an ISM which had a small or null enrichment 
during  the second half of the Universe lifetime. In this scenario, their low mass 
(see $\S$4.3) is in agreement with no nitrogen dredge-up, and thus low N/O.


\section{The star formation history of IC10}

The spectroscopically observed PNe,   all with low N/O ratio and/or 
 with very low or absent \heii, are, in a 
first approximation, \lq old stars', thus they  could be born in a
limited period of time, 7 to 11 Gyr ago, i.e., during the first half of 
the age of the Universe.  Following the discussion
of \citet{magrini05}, we obtain that the mass of the stellar 
population born during the epoch of the PN progenitors' formation, is
proportional to the number of PNe. About 2$\times$10$^6$M$_{\odot}$
were formed by each observed PN.  

Thus, we considered the two surveys for PNe in IC10: the present one and that by 
\citet{magrini03}.  The current survey is the deeper one, since it discovered 12 PNe 
in an area of 5.5\arcmin$\times$5.5\arcmin, where the previous survey found only three. 
The previous study surveyed an area of 33\arcmin$\times$33\arcmin\ and lead to the discovery 
of 15 candidate PNe.  If the survey by \citet{magrini03} reached the same depth as the present 
search, this would imply  the presence of about 60 (i.e., four times the previous number of known PNe).
Thus  the mass of the underlying population would be approximately
1.2$\times$10$^8$M$_{\odot}$.  This corresponds to a star formation rate
(SFR) of about 0.02 M$_{\odot}$ yr$^{-1}$ if the star formation bursts were 
extended through
a long period of about 7 Gyr. If we consider
that all PNe belong to the same age range, 7-10 Gyr old, thus the SFR
would be 0.04 M$_{\odot}$ yr$^{-1}$.  This SFR is comparable to that
found at present time using the observed \ha\ flux: 0.04-0.08
M$_{\odot}$ yr$^{-1}$ \citep{thronson90}.

\section{The metal distribution in the ISM of IC10}

Recent observations allowed a deeper knowledge of nearby dwarf galaxies, 
resolving the spatial distribution of their metal content. These results  
are giving indications of a non-homogeneous chemical composition also at large 
scales, as, for instance,  
in the dwarf irregular galaxies of the Local Group and nearby groups NGC~6822 
\citep{Lea06}, Sextans~B (\citealt{kniazev05}, \citealt{Mea05}), NGC~4214 
(\citealt{drosdo02}, \citealt{kolbul96}); and in the blue compact galaxies  
SBS0335-052 \citep{Izotov06} and Mrk86 \citep{gildepaz99}. 

Our and previous observations of IC10 show evidence of abundance 
non-homogeneities. The \hii\ regions (and PNe)
under analysis in this paper are located in the central
5.5\arcmin$\times$5.5\arcmin, that, at the distance to IC10, means a
projected area of 1.3$\times$1.3 kpc. Typical angular distances
between the \hii\ regions of our sample are $\sim$1-1.5\arcmin, thus
$\sim$300-400 kpc.  The average abundance of oxygen in \hii\ regions,
considering only those where at least one electron temperature,
\teoiii\ and/or \teoii\ was measured, is 2.2$\pm$1.3$\times$10$^{-4}$
by number, or 12+$\log$(O/H)=8.30$\pm$0.20. 
The dispersion is comparable with those of other dwarf irregular galaxies 
whose metallicity map was recently obtained. For instance, the average 
oxygen abundances, are: 7.77 with a smaller dispersion of 0.07 dex for
the \hii\ regions of NGC~3109 \citep{pena07}, 7.6$\pm$0.2 for Sextans A, 
and 7.8$\pm$0.2 for Sextans B (\citealt{kniazev05}, \citealt{Mea05}); 
and 8.06$\pm$0.09 for the SMC (see \citealt{pena07}).

Chemical abundance determinations of \hii\ regions in IC10 were first 
derived by \citet{lequeux79} who observed its two brightest regions, 
named IC10-1 and IC10-2, which we re-identified as HL90-45 and HL90-111 
\citep{hodge90}.  
The study of Lequeux and collaborators indicated a notable difference in 
the chemical compositions of \hii\ regions located in different areas of 
the galaxy: HL90-45 was found to have 12+$\log$(O/H)=8.45 whereas for HL90-111
this abundance is 8.11, in good agreement with our results shown in
Table~5. More recently, \citet{richer01} obtained chemical
abundances of three IC10 \hii\ regions: HL90-106b (with O/H
7.86$\pm$0.32), HL90-111c (7.84$\pm$0.25), HL90-111d (8.23$\pm$0.09).
Again a notable difference in chemical composition is found and, in addition,
a difference in the chemical composition within the same \hii\ region 
complex, HL90-111, was derived: HL90-111c and Hl90-111d show that the
variance of chemical abundances is in place also at small scales. The
inhomogeneity revealed in HL90-111 is similar to that found in 30
Doradus by \citet{tsamis05} for which they claimed the evidence for
incomplete small-scale mixing of the ISM.
In addition, very recent deep optical data collected by \citet{sanna09} 
suggest the presence of a spread in heavy 
element abundances of the order of one-half dex. 

The O/H abundance vs. galactocentric radius is shown in Figure 3. 
Due to the uncertainties on the position angle and on the inclination 
of IC10, the galactocentric radius has been computed without any de-projection, 
thus considering that at a distance of 750 kpc, 1\arcsec\ corresponds to 3.6 pc. 
Note that there is no indication of radial metallicity gradients, neither 
for \hii\ regions nor for PNe.

The scale-length of metallicity non-homogeneities is of the order of 0.2-0.3 kpc, 
with clumps showing the higher metallicities,  HL90-29 and HL90-45, and  
an area with a more constant metallicity, around 12+$\log$O/H=8.2, located  
between 0.2 to 0.5 kpc from the center. For PNe the abundance determination are 
upper limits, with the exception of PN17, thus their metallicity map remains quite 
uncertain.

\begin{figure} 
 \centering
 \includegraphics[width=8truecm]{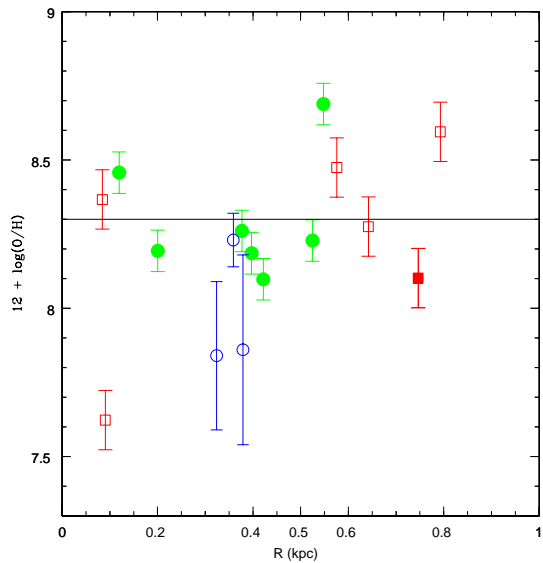} 
 \caption{Oxygen abundance vs. galactocentric radius, not de-projected.
 Symbols represents data from: \hii\ regions from the present paper, only those with \te\ measured (filled circles); \hii\ 
 regions from \citet{richer01} (empty circles); PNe from the present work (filled square with \te\ measured, empty square 
 with upper limit of \te). The continuous line represents the average O/H from our \hii\ regions.  }
 \label{homo}
\end{figure}

\section{The third dredge-up and its effect on the  PN composition}

In contrast to \hii\ regions, some elemental abundances in PNe are
affected by the nucleosynthesis in the PN progenitors. Newly
synthesized material can be dredged up by convection in the envelope,
significantly altering the abundances of He, C and N in the surface layers
during the evolution of the PN progenitor stars on the giant branch
and asymptotic giant branch (AGB). Also a certain amount of oxygen can
be mixed in during the thermally pulsing phase of the AGB evolution
(\citealt{KB94}; \citealt{pequignot00}; \citealt{leisy06}).

However, the dredge-up of oxygen and neon seems a rare event,
happening mainly in low metallicity environments. This has induced us to
believe that oxygen is a good tracer of the ISM abundances at the
epoch of the PN progenitor formation. Recent studies have
analyzed the circumstances that make the dredge-up of oxygen possible,
both from theoretical and observational points of view.

\citet{pequignot00} argue that oxygen is a by-product of all third
dredge-up, but leads to enrichment only at low metallicity. At solar
metallicity, the dredged-up material has lower oxygen abundance than
the original gas. \citet{RM07} analyzed the abundances for a sample of
bright PNe in nearby dwarf irregular galaxies. They suggest that the
dredge-up of oxygen is an infrequent phenomenon also in low
metallicity environment. \citet{Mea05} and \citet{kniazev05} compared
the chemical abundances of \hii\ regions of the dwarf galaxy Sextans A
with that of the PN known in this galaxy. Both authors found
significant self-pollution of the PN progenitor, by a factor of
$\sim$10 in oxygen. \citet{leisy06} found several examples of oxygen
enrichment in the PNe of SMC, while their number was reduced in the LMC, 
which has a higher metallicity.  \citet{kniazev07} found an enrichment
in oxygen by 0.27$\pm$0.10 dex in the PN of the Fornax galaxy respect
to the ISM. Finally, \citet{kniazev08} found that only one of the five
PNe in the dwarf galaxy Sagittarius shows the effect of the third
dredge-up. They argue that the different behavior respect to the
remaining PN population of Sagittarius could be due to a different
characteristic of the progenitor, such as the presence of rotation
\citep{Siess04} and/or a different age (and thus mass).

\begin{figure}
\centering
\includegraphics[width=8.5truecm]{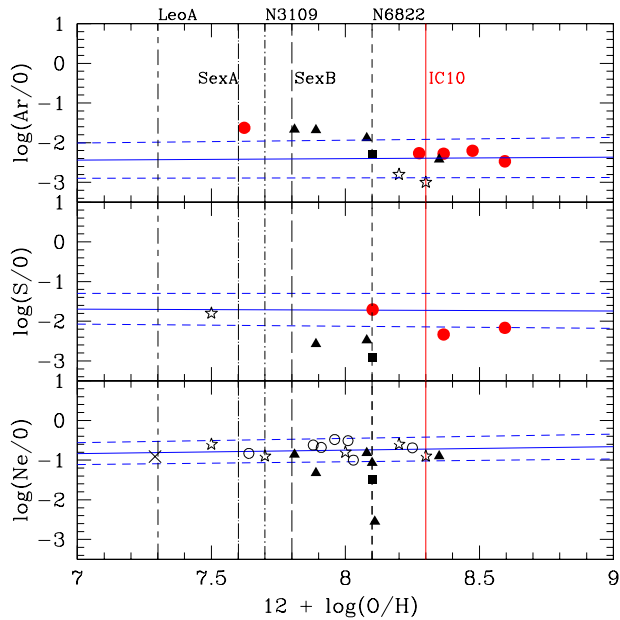} 
\caption{The planetary nebulae Ne/O, S/O and Ar/O abundance ratios as a function of the oxygen abundance 
of the PNe in a number of nearby dwarf irregular galaxies. Symbols for IC10 PNe are filled circles. 
For the other galaxies they are as follows:  cross for Leo~A; filled triangles for NGC~3109; filled square
for Sex~A; empty stars for Sex~B; and empty circles for NGC~6822. The quasi-horizontal lines show the abundance 
line ratios given by the a sample \hii\ regions of blue compact galaxies, as compiled by \citet{Izotov06}. 
The vertical lines show the 12+log(O/H) of the \hii\ regions of each of the dIrr in this plot. These lines 
are labeled with the corresponding galaxy identification at the top of the plot.}
\label{3du}
\end{figure}

In Figure~4 we study the presence or not of the third-dredge in a
number of nearby dwarf irregular galaxies (or dwarf irregulars,
dIrs). We do not plot the data for dwarf spheroidal galaxies since the
comparison with the present-time ISM is not possible.  The idea behind
the comparison of PNe abundance ratios as a function of the oxygen
abundance in different galaxies, with the 12+log(O/H) of the
\hii\ regions in each of these galaxies (vertical lines), and in a
sample of BCGs (\lq horizontal' lines), is that in \hii\ regions, the
oxygen that is seen has been produced by the same massive stars that
produced the alpha-process neon, sulfur and argon. Therefore, in this
plot, log(Ne/O), log(S/O) and log(Ar/O) should be constant and show no
dependence on the oxygen abundance. In fact, inspecting Figure~4 we
see that in the three diagrams most of the PNe are distributed along
the line that defines the BCGs abundance ratios of \hii\ regions, thus
with no trends with respect to the oxygen abundance.
The oxygen production in these diagrams should have
two effects: {\em i)} to place the PN points well ahead
of the mean abundance in the ISM, {\em ii)}  to depress the Ne/O,
S/O, and Ar/O ratios, unless the production of Ne, S, and Ar also
occurs. 
In a conservative view, we would aspect that the presence of both effects would 
probe the third-dredge-up occurrence. 
Few PNe pass this test full-filling both requests, usually
with abundance ratios that are not completely understood. Note 
from the plot that for galaxies of much lower metallicity this contribution 
is important, as in the case of Sex~A [12+log(O/H)=7.6] and NGC~3109
[12+log(O/H)=7.7] (see Ne/O and S/O, but also their displacement from the average ISM O/H). 
In fact, previous authors, when discussing the abundance patterns of the PN 
population of these two galaxies, have already pointed out this effect 
(\citealt{Mea05} and \citealt{pena07}, respectively).
Being less conservative, also some  PNe of Sextans B, whose O/H abundances are ahead of 
the mean abundance in the ISM, could be affected by oxygen production. 
However, at the metallicity of IC10 [12+log(O/H)=8.3] we do not find 
important contributions of dredge-up in the abundance of oxygen: all PNe have  O/H
close to the present-time ISM one, and no particularly depressed Ne/O, Ar/O, and S/O ratios are observed. 

Remembering, as discussed above, that the third dredge-up of oxygen is
a metallicity dependent phenomenon, this kind of plots can
also be useful to give a lower limit above of which no third dredge-up
effect is seen in the nearby dwarf irregulars. So far, as determined by
all the available data in the literature, this limit is around
12+log(O/H)=7.7.

\section{Summary and conclusions}
In this paper we present imaging and spectroscopic observations of a
sample of PNe and \hii\ regions in the nearby starburst galaxy
IC10. The pre-imaging observations allowed us to discover 12 new PNe in
the central 5.5\arcmin$\times$5.5\arcmin\ regions. We derived physical
and chemical properties of 6 PNe (one of them with direct electron
temperature measurement) and of 10 \hii\ regions (7 with direct
electron temperature). Using the PN \ha\ flux and the upper limit to
the \heii/\hb\ ratio we evaluated the mass and the formation epoch 
of the PN progenitors.  From the analysis of the above properties of
PNe and \hii\ regions we found:
\begin{itemize} 
\item in our rough age estimation, PN progenitors are coeval within few Gyr and were born between 7 and 10 Gyr ago;
\item the mean oxygen abundances of PNe and \hii\ regions are similar, indicating a small chemical enrichment;
\item both from PNe and \hii\ regions, the mean oxygen abundance is sub-solar, 12+$\log$(O/H)=8.3; 
\item the metallicity of the interstellar medium was not homogeneous, neither at present-time (from \hii\ 
regions) nor in the past (from PNe);
\item the third-dredge up of oxygen does not occur in the PNe of IC10, and from a comparative analysis of PNe in 
dwarf irregular galaxies it seems to be allowed, but not mandatory,  
for 12+$\log$(O/H)$\leq$7.7.  
\end{itemize}

\section*{Acknowledgments} 
We thank Michael Richer, the referee of this paper, for his careful reading of the manuscript 
and for his comments which improved the quality and the presentation of the paper.  
Two Brazilian and one Italian agency gave us partial support for this work. 
So DRG and LM would like to thank FAPESP for its grants: 2003/09692-0 and 2006/59301-6, 
respectively, and the FAPERJ's grant, E-26/110.107/2008.

\appendix
\section{Emission-line flux measurements}
In this section we present the observed emission-line fluxes and extinction corrected intensities 
measured in the samples of PNe and \hii\ regions of IC10.

\begin{table}
\centering
\begin{minipage}{85mm}
{\scriptsize  
\caption{Observed and extinction corrected fluxes of PNe. Column
(1) gives the PN name; column (2) gives the observed \hb\ flux 
in units of 10$^{-17}$erg cm$^{-2}$ s$^{-1}$; column (3) the nebular extinction
coefficient; columns (4) and (5) indicate the emitting ion and the rest frame wavelength
in \AA; columns (6), (7), and (8) give the measured (F$_{\lambda}$), the relative error on the 
measured fluxes ($\Delta$F$_{\lambda}$) and the extinction
corrected (I$_{\lambda}$) intensities. Both F$_{\lambda}$
and I$_{\lambda}$ are normalized to \hb=100. Upper limits on the line fluxes are marked with $:$.}
\begin{tabular}{@{}ccclcrrr@{}}
\hline
Id & F$_{{\rm H}\beta}$ & c({H$\beta$}) & Ion & $\lambda$ (\AA) & F$_{\lambda}$ & $\Delta$F$_{\lambda}$& I$_{\lambda}$ \\ 
\hline
IC10PN5 & 1.3  & 1.95 & [OIII]  & 4363  & 9:	   & -     & 16.7\\ 
        &      &      & HeII	& 4686  & 9:	   & -     & 10.7\\ 
        &      &      & HI	& 4861  & 100.0    & 08\%  & 100.\\ 
        &      &      & [OIII]  & 4959  & 441.0    & 06\%  & 395.\\ 
        &      &      & [OIII]  & 5007  & 1384.    & 05\%  & 1179.\\ 
        &      &      & HeI	& 5876  & 47.0    & 10\%  & 18.6\\ 
        &      &      & [NII]	& 6548  & 43.5    & 10\%  & 11.2\\ 
        &      &      & HI	& 6563  & 1113.    & 06\%  & 285.\\ 
        &      &      & [NII]	& 6584  & 129.0    & 08\%  & 32.6\\ 
        &      &      & HeI	& 6678  & 16.7    & 12\%  & 4.0\\ 
        &      &      & [SII]	& 6717  & 15.3    & 12\%  & 3.5\\ 
        &      &      & [SII]	& 6731  & 13.2    & 12\%  & 3.0\\ 
        &      &      & HeI	& 7065  & 35.1    & 10\%  & 6.6\\ 
        &      &      & [ArIII] & 7135  & 59.4    & 9.5\% & 10.6\\ 
        &      &      & [SIII]  & 9069  & 65.6    & 9.0\% & 3.5\\ 
\hline
IC10PN9& 3.7  & 1.49& [OIII]  & 4363  & 9:	&  -	 &  9:   \\	
       &      &     & HeII    & 4686  & 3:	&  -	 &  3:   \\
       &      &     & HI      & 4861  & 100.	&  08\%  &  100. \\
       &      &     & [OIII]  & 4959  & 521.	&  05\%  &  482. \\
       &      &     & [OIII]  & 5007  & 1551.	&  04\%  &  1375.\\
       &      &     & [NII]   & 6548  & 33.	&  09\%  &  12.  \\
       &      &     & HI      & 6563  & 809.	&  05\%  &  285. \\
       &      &     & [NII]   & 6584  & 98.7	&  08\%  &  34.5 \\
       &      &     & HeI     & 6678  & 13.53	&  10\%  &  4.54 \\	 
       &      &     & [SII]   & 6717  & 6.3	&  11\%  &  2.0  \\
       &      &     & [SII]   & 6731  & 11.9	&  10\%  &  3.9  \\
       &      &     & HeI     & 7065  & 39.49	&  09\%  &  10.9 \\	 
       &      &     & [ArIII] & 7135  & 38.9	&  09\%  &  10.4 \\
       &      &     & [OII]   & 7320  & 24.81	&  10\%  &  6.1  \\
       &      &     & [OII]   & 7330  & 36.0	&  09\%  &  8.8  \\
\hline
PN17 & 6.4  & 1.18& HI        & 4340  & 19.0     & 26\%  &  27.2 \\ 
     &      &     & [OIII]    & 4363  & 5.5       & 28\% &  7.7 \\     
     &      &     & HeII      & 4686  & 7.3      & -      &  8.2   \\    
     &      &     & HI        & 4861  & 100.00   & 16\% &  100.  \\ 
     &      &     & [OIII]    & 4959  & 166.2   & 14\% &  155. \\    
     &      &     & [OIII]    & 5007  & 473.4   & 09\% &  429. \\     
     &      &     & HeI       & 5876  & 29.3    & 22\% &  16.7\\     
     &      &     & [NII]     & 6548  & 31.0    & 22\% &  13.6\\     
     &      &     & HI        & 6563  & 653.   & 12\%  &  285. \\
     &      &     & [NII]     & 6584  & 96.7    & 22\% &  41.9 \\      
     &      &     & [SII]     & 6731  & 13.8    & 28\% &  5.8   \\  
     &      &     & HeI       & 6678  & 7.6     & 26\% &  3.1  \\     
     &      &     & HeI       & 7065  & 15.8    & 23\% &  5.7  \\     
     &      &     & [OII]     & 7320  & 23.5    & 23\% &  7.7  \\     
     &      &     & [OII]     & 7330  & 20.5    & 22\% &  6.7  \\     
     &      &     & [SIII]    & 9069  & 30.3    & 22\% &  8.2 \\           
\hline
PN20 & 2.1 & 1.89 & [OIII]   & 4363  & 5:     &  -    &   9:  \\
     &      &      & HI        & 4861  & 100.0 & 10\%  & 100.0\\
     &      &      & [OIII]    & 4959  & 409. & 09\%  & 368.\\ 
     &      &      & [OIII]    & 5007  & 1236.& 06\%  & 1061.\\  
     &      &      & HeI       & 5876  & 24.6  & 13\%  & 10.0 \\
     &      &      & [NII]     & 6548  & 58.3  & 12\%  & 15.7 \\
     &      &      & HI        & 6563  & 1070.& 06\%  & 285.\\
     &      &      & [NII]     & 6584  & 177. & 9.5\% & 46.5 \\ 
     &      &      & HeI       & 6678  & 21.  & 13\%  & 5.3  \\   
     &      &      & [SII]     & 6717  & 33.4  & 13\%  & 8.1  \\
     &      &      & [SII]     & 6731  & 53.0  & 12\%  & 12.8 \\
     &      &      & HeI       & 7065  & 32.9  & 13\%  & 6.7  \\
     &      &      & [ArIII]   & 7135  & 68.7  & 11\%  & 12.9 \\ 
     &      &      & [SIII]    & 9069  & 53.8  & 12\%  & 18.3 \\
\hline
\end{tabular}
}
\end{minipage}
\label{tabPN_flux}
\end{table}
\begin{table}
\centering
\begin{minipage}{85mm}
{\scriptsize  
\contcaption{}
\begin{tabular}{@{}ccclcrrr@{}}
\hline
Id & F$_{{\rm H}\beta}$ & c({H$\beta$}) & Ion & $\lambda$ (\AA) & F$_{\lambda}$ & $\Delta$F$_{\lambda}$ & I$_{\lambda}$ \\ 
\hline
PN21 & 3.1 & 1.06& [OIII]   & 4363  & 6:     & -    &  7.8   \\
     &      &     & HI       & 4861  & 100. & 22\% &  100. \\
     &      &     & [OIII]   & 4959  & 141. & 22\% &  133. \\
     &      &     & [OIII]   & 5007  & 539. & 19\% &  496. \\
     &  &  & [NII]     & 6548  & 88.5  & 22\% &  42.  \\
     &  &  & HI        & 6563  & 599. & 18\% &  285. \\
     &  &  & [NII]     & 6584  & 275. & 20\% &  130. \\
     &  &  & HeI       & 6678  & 14.2  & 23\% &  8.6  \\
     &  &  & [SII]     & 6717  & 26.4  & 23\% &  11.9  \\
     &  &  & [SII]     & 6731  & 39.3  & 23\% &  17.7  \\
     &  &  & [ArIII]   & 7135  & 29.2  & 23\% &  11.5  \\
     &  &  & [SIII]    & 9069  & 49.4  & 22\% &  10.0  \\    
\hline
PN22 & 3.5 & 1.65& [OII]   & 3727  & 198. & 12\% & 577.\\
     &     &	 & [OIII]    & 4363  & 14:    &  -   & 22:   \\
     &     &	 & HeII      & 4686  & 10:   &  -   & 12: \\
     &     &	 & HI	     & 4861  & 100.& 13\% & 100.\\
     &     &	 & [OIII]    & 4959  & 445. & 10\% & 409.\\
     &     &	 & [OIII]    & 5007  & 1415. & 06\% & 1241.\\
     &     &	 & HeI      & 5876  & 25.     & 06\% & 11.\\
     &     &	 & [NII]     & 6548  & 61.7  & 13\% & 19.6 \\
     &     &	 & HI	     & 6563  & 904. & 06\% & 285.\\
     &     &	 & [NII]     & 6584  & 162. & 12\% & 50.5 \\
     &     &	 & [SII]     & 6717  & 9.4   & 21\% & 2.7  \\
     &     &	 & [SII]     & 6731  & 15.5  & 20\% & 4.5  \\
     &     &	 & HeI       & 7065  & 42.6  & 16\% & 10.3 \\
     &     &	 & [ArIII]   & 7135  & 67.2  & 13\% & 15.6 \\ 
     &     &	 & [SIII]    & 9069  & 80.2  & 13\% & 6.7  \\  
\hline
PN23 & 0.55& 1.48& HeII    & 4686  & 19:     & -    & 23:  \\
     &     &	 & HI	     & 4861  & 100.  & 24\% & 100. \\ 
     &     &	 & [OIII]    & 4959  & 69.6   & 28\% & 64.  \\
     &     &	 & [OIII]    & 5007  & 164.  & 20\% & 145.  \\
     &     &	 & [NII]     & 6548  & 113.  & 20\% & 41.  \\
     &     &	 & HI	     & 6563  & 801.  & 10\% & 285. \\
     &     &	 & [NII]     & 6584  & 341.  & 17\% & 120.  \\
     &     &	 & [SII]     & 6717  & 46.2   & 28\% & 15.3  \\
     &     &	 & [SII]     & 6731  & 36.1   & 30\% & 11.9  \\
     &     &	 & HeI       & 7065  & 9.8    & 36\% & 2.8   \\
\hline
PN24 & 3.7 & 1.41& HI	     & 4861  & 100.  & 09\% & 100.\\
     &     &	 & [OIII]    & 4959  & 611.  & 06\% & 565.\\
     &     &	 & [OIII]    & 5007  & 1780. & 05\% & 1601.\\
     &     &	 & HeI       & 5876  & 57.   & 09\% & 29. \\
     &     &	 & [NII]     & 6548  & 60.   & 09\% & 22.5 \\
     &     &	 & HI	     & 6563  & 764.  & 5.5\%& 285.\\
     &     &	 & [NII]     & 6584  & 169.  & 07\% & 62.5 \\
     &     &	 & [SII]     & 6731  & 32.1   & 9.5\%& 11.1 \\
     &     &	 & [ArIII]   & 7135  & 59.9   & 09\% & 17.2 \\
\hline
PN25 & 0.22 & 0.92 & HI        & 4861  & 100. & 12\% & 100.  \\
     &      &	   & [OIII]    & 4959  & 386. & 10\% & 367. \\    
     &      &	   & [OIII]    & 5007  & 1140.& 06\% & 1058. \\  
     &      &	   & HI        & 6563  & 544. & 11\% & 285.  \\ 
\hline
\end{tabular}
}
\end{minipage}
\label{tabPN_flux}
\end{table}

\begin{table}
\centering
\begin{minipage}{85mm}
{\scriptsize  
\caption{Observed and de-reddened fluxes of \hii\ regions. Column (1) gives the name of the \hii\ region. The 
rest of the columns are as in Table~2.}
\begin{tabular}{@{}ccclcrrr@{}}
\hline
ID & F$_{{\rm H}\beta}$ & c({H$\beta$}) & Ion & $\lambda$ (\AA) & F$_{\lambda}$ & $\Delta$F$_{\lambda}$ & I$_{\lambda}$ \\
\hline
IC10PN7& 111 & 1.63& [OII]     & 3727  & 35.45 & 07\%  &  101. \\ 
       &     &     & [NeIII]   & 3869  & 5.05  & 11\%  &  12.7 \\    
       &     &     & [NeIII]/HI& 3968  & 7.53  & 10\%  &  17.2 \\   
       &     &     & HI        & 4100  & 10.4  & 09\%  &  21.1 \\
       &     &     & HI        & 4340  & 27.8  & 6.5\% &  45.2 \\
       &     &     & [OIII]    & 4363  & 1.9:  & -     &  3.5: \\   
       &     &     & HI        & 4861  & 100.  & 05\%  &  100. \\
       &     &     & [OIII]    & 4959  & 148.  & 4.5\% &  135. \\	  
       &     &     & [OIII]    & 5007  & 455.  & 03\%  &  397. \\  
       &     &     & HeI       & 5876  & 25.3  & 6.5\% &  11.7 \\
       &     &     & [OI]      & 6300  & 3.74  & 10\%  &  1.38 \\
       &     &     & [OI]      & 6363  & 3.22  & 10\%  &  1.15 \\
       &     &     & [NII]     & 6548  & 25.2  & 6.5\% &  8.15 \\
       &     &     & HI        & 6563  & 889.  & 05\%  &  285. \\
       &     &     & [NII]     & 6584  & 76.2  & 03\%  &  24.1 \\  
       &     &     & HeI       & 6678  & 13.2  & 05\%  &  4.01 \\
       &     &     & [SII]     & 6717  & 39.8  & 9.5\% &  11.8 \\
       &     &     & [SII]     & 6731  & 32.1  & 6.5\% &  9.43 \\
       &     &     & HeI       & 7065  & 9.42  & 6.5\% &  2.32 \\
       &     &     & [ArIII]   & 7135  & 48.5  & 6.5\% &  11.5 \\
       &     &     & [OII]     & 7320  & 8.72  & 09\%  &  1.88 \\
       &     &     & [OII]     & 7330  & 6.98  & 09\%  &  1.50 \\
       &     &     & [ArIII]   & 7751  & 11.9  & 9.5\% &  2.07 \\
       &     &     & [SIII]    & 9069  & 67.2  & 06\%  &  5.81 \\ 
\hline
HL90-13& -   & 0.04& HI        & 4861  & 100. & 05\% & 100. \\
       &     &     & [OIII]    & 4959  & 110. & 05\% & 110. \\
       &     &     & [OIII]    & 5007  & 327. & 04\% & 326. \\
       &     &     & HeI       & 5876  & 9.53 & 08\% & 9.36 \\
       &     &     & [NII]     & 6548  & 7.99 & 08\% & 7.78 \\
       &     &     & HI        & 6563  & 292. & 04\% & 285. \\
       &     &     & [NII]     & 6584  & 25.1 & 07\% & 24.5 \\  
       &     &     & HeI       & 6678  & 3.68 & 09\% & 3.68 \\  
       &     &     & [SII]     & 6717  & 20.4 & 07\% & 19.8 \\  
       &     &     & [SII]     & 6731  & 16.0 & 07\% & 15.5 \\  
       &     &     & [ArIII]   & 7135  & 20.7 & 07\% & 20.1 \\
       &     &     & [OII]     & 7320  & 5.67 & 09\% & 5.48 \\ 
       &     &     & [OII]     & 7330  & 3.99 & 09\% & 3.85 \\
       &     &     & [ArIII]   & 7751  & 5.74 & 09\% & 5.51 \\
       &     &     & [SIII]    & 9069  & 36.0 & 06\% & 34.0 \\  
\hline
HL90-20& -   & 1.34& [OII]     & 3727  & 42.1 &  09\%   & 100. \\
       &     &     & [NeIII]   & 3968  & 6.04 &  14.5\% & 12.0 \\
       &     &     & HI        & 4100  & 13.5 &  14.5\% & 24.3 \\
       &     &     & HI        & 4340  & 33.0 &  09\%   & 49.4 \\
       &     &     & [OIII]    & 4363  & 2.71:&  -	& 4.20:\\
       &     &     & HeI       & 4471  & 3.12 &  15\%   & 4.21 \\
       &     &     & HI        & 4861  & 100. &  08\%   & 100. \\
       &     &     & [OIII]    & 4959  & 118. &  08\%   & 110. \\
       &     &     & [OIII]    & 5007  & 379. &  06\%   & 341. \\
       &     &     & HeI       & 5876  & 24.8 &  10\%   & 13.1 \\
       &     &     & [NII]     & 6548  & 15.3 &  14.5\% & 6.05 \\
       &     &     & HI        & 6563  & 726. &  05\%   & 285. \\
       &     &     & [NII]     & 6584  & 41.5 &  09\%   & 16.1 \\
       &     &     & HeI       & 6678  & 11.8 &  14.5\% & 4.47 \\ 
       &     &     & [SII]     & 6717  & 26.9 &  10\%   & 9.91 \\
       &     &     & [SII]     & 6731  & 20.8 &  11\%   & 7.61 \\
       &     &     & HeI       & 7065  & 7.85 &  15\%   & 2.48 \\
       &     &     & [ArIII]   & 7135  & 36.5 &  09\%   & 11.2 \\
       &     &     & [OII]     & 7320  & 5.05 &  15\%   & 1.43 \\
       &     &     & [OII]     & 7330  & 3.82 &  15\%   & 1.08 \\
       &     &     & [ArIII]   & 7751  & 9.42 &  14\%   & 2.23 \\
       &     &     & [SIII]    & 9069  & 73.3 &  07\%   & 9.86 \\ 
\hline
\end{tabular}
}
\end{minipage}
\label{tabHII_flux}
\end{table}

\begin{table}
\centering
\begin{minipage}{85mm}
{\scriptsize  
\contcaption{}
\begin{tabular}{@{}ccclcrrr@{}}
\hline
ID & F$_{{\rm H}\beta}$ & c({H$\beta$}) & Ion & $\lambda$ (\AA) & F$_{\lambda}$ & $\Delta$F$_{\lambda}$ & I$_{\lambda}$ \\ 
\hline
HL90~29& 128 & 1.22& [OII]     & 3727  & 110. & 13\% &  244. \\ 
       &     &     & HI        & 4340  & 36.1 & 19\% &  52.2 \\
       &     &     & [OIII]    & 4363  & 8.33 & 28\% &  14.5 \\
       &     &     & HI        & 4861  & 100. & 13\% &  100. \\
       &     &     & [OIII]    & 4959  & 254. & 09\% &  238. \\
       &     &     & [OIII]    & 5007  & 742. & 06\% &  673. \\
       &     &     & HeI       & 5876  & 32.1 & 19\% &  17.9 \\
       &     &     & [NII]     & 6548  & 40.4 & 18\% &  17.3 \\
       &     &     & HI        & 6563  & 670. & 06\% &  285. \\
       &     &     & [NII]     & 6584  & 90.4 & 12\% &  38.1 \\
       &     &     & [SII]     & 6717  & 99.0 & 11\% &  39.6 \\
       &     &     & [SII]     & 6731  & 74.0 & 13\% &  29.4 \\
       &     &     & [ArIII]   & 7135  & 56.1 & 17\% &  19.1 \\
       &     &     & [OII]     & 7320  & 9.20 & 23\% &  2.91 \\
       &     &     & [OII]     & 7330  & 8.74 & 23\% &  2.74 \\
       &     &     & [ArIII]   & 7751  & 13.6 & 21\% &  3.64 \\
       &     &     & [SIII]    & 9069  & 49.0 & 13\% &  11.8 \\
\hline
HL90-30& 110 & 1.63& HI        & 4340  & 28.5 &  18\% &  46.5 \\
       &     &     & [OIII]    & 4363  & 3.00:&  -    &  4.8: \\
       &     &     & HI        & 4861  & 100. &  14\% &  100. \\
       &     &     & [OIII]    & 4959  & 116. &  14\% &  106. \\
       &     &     & [OIII]    & 5007  & 407. &  13\% &  357. \\
       &     &     & HeI       & 5876  & 29.5 &  18\% &  13.5 \\
       &     &     & [NII]     & 6548  & 25.1 &  18\% &  8.13 \\ 
       &     &     & HI        & 6563  & 894. &  12\% &  286. \\
       &     &     & [NII]     & 6584  & 65.4 &  14\% &  20.7 \\ 
       &     &     & HeI       & 6678  & 12.7 &  17\% &  3.84 \\
       &     &     & [SII]     & 6717  & 48.9 &  15\% &  14.4 \\ 
       &     &     & [SII]     & 6731  & 39.6 &  15\% &  11.6 \\ 
       &     &     & [ArIII]   & 7135  & 49.1 &  15\% &  11.6 \\
       &     &     & [OII]     & 7320  & 9.57 &  17\% &  2.06 \\
       &     &     & [OII]     & 7330  & 8.04 &  17\% &  1.72 \\
       &     &     & [ArIII]   & 7751  & 11.5 &  17\% &  1.98 \\
       &     &     & [SIII]    & 9069  & 81.4 &  14\% &  6.98 \\ 
\hline 
HL90-40& 10.3& 1.43& [OII]     & 3727  & 106. &  05\% &	270. \\ 
       &     &     & HI        & 4340  & 22.7 &  17\% & 35.0 \\
       &     &     & HI        & 4861  & 100. &  05\% & 100. \\
       &     &     & [OIII]    & 4959  & 50.5 &  14\% & 46.9 \\
       &     &     & [OIII]    & 5007  & 153.6&   09\%& 137. \\
       &     &     & HeI       & 5876  & 24.0 &  17\% & 12.1 \\
       &     &     & [OI]      & 6300  & 9.17 &  20\% & 3.82 \\ 
       &     &     & [NII]     & 6548  & 38.8 &  15\% & 14.4 \\
       &     &     & HI        & 6563  & 777. &  04\% & 286. \\
       &     &     & [NII]     & 6584  & 110. &  06\% & 40.1 \\ 
       &     &     & HeI       & 6678  & 13.3 &  20\% & 4.66 \\ 
       &     &     & [SII]     & 6717  & 142. &  05\% & 48.7 \\ 
       &     &     & [SII]     & 6731  &  98.1&  05\% & 33.4 \\ 
       &     &     & HeI       & 7065  & 5.42 &  23\% & 1.58 \\
       &     &     & [ArIII]   & 7135  & 25.2 &  17\% & 7.14 \\
       &     &     & [SIII]    & 9069  & 41.7 &  15\% & 4.84 \\  
\hline
\end{tabular}
}
\end{minipage}
\label{tabHII_flux}
\end{table}

\begin{table}
\centering
\begin{minipage}{85mm}
{\scriptsize  
\contcaption{}
\begin{tabular}{@{}ccclcrrr@{}}
\hline
ID & F$_{{\rm H}\beta}$ & c({H$\beta$}) & Ion & $\lambda$ (\AA) & F$_{\lambda}$ & $\Delta$F$_{\lambda}$ & I$_{\lambda}$ \\ 
\hline
HL90-45       & 1025.& 1.81 & [OII]     & 3727  & 31.0 & 26\%  & 100. \\ 
(BR1)         &      &      & HI        & 3835  & 2.30 & 30\%  & 6.69 \\
    	      &      &      & [NeIII]   & 3869  & 16.6 & 26\%  & 46.6 \\
    	      &      &      & HeI       & 3889  & 6.67 & 30\%  & 18.3 \\
    	      &      &      & [NeIII]/HI& 3968  & 12.0 & 26\%  & 30.4 \\
    	      &      &      & HI        & 4100  & 12.4 & 26\%  & 27.5 \\
    	      &      &      & HI        & 4340  & 30.8 & 24\%  & 53.1 \\
    	      &      &      & [OIII]    & 4363  & 2.84 & 30\%  & 4.78 \\
    	      &      &      & HeI       & 4471  & 3.17 & 30\%  & 4.76 \\
    	      &      &      & HeI/[ArIV]& 4712  & 0.79 & 35\%  & 0.93 \\
    	      &      &      & [ArIV]    & 4740  & 0.24 & 35\%  & 0.27 \\
    	      &      &      & HI        & 4861  & 100. & 19\%  & 100. \\
    	      &      &      & [OIII]    & 4959  & 275. & 13\%  & 251. \\
    	      &      &      & [OIII]    & 5007  & 874. & 11\%  & 757. \\
    	      &      &      & [NII]     & 5755  & 0.32 & 35\%  & 0.14 \\
    	      &      &      & HeI       & 5876  & 33.7 & 24\%  & 14.2 \\
    	      &      &      & [OI]      & 6300  & 3.33 & 30\%  & 1.09 \\
    	      &      &      & [OI]      & 6363  & 0.85 & 35\%  & 0.27 \\
    	      &      &      & [NII]     & 6548  & 10.0 & 26\%  & 2.86 \\
    	      &      &      & HI        & 6563  &1012. & 11\%  & 285. \\
    	      &      &      & [NII]     & 6584  & 35.5 & 24\%  & 9.89 \\ 
    	      &      &      & HeI       & 6678  & 15.1 & 26\%  & 4.02 \\ 
    	      &      &      & [SII]     & 6717  & 25.2 & 25\%  & 6.50 \\
    	      &      &      & [SII]     & 6731  & 24.7 & 25\%  & 6.33 \\ 
    	      &      &      & HeI       & 7065  & 20.6 & 26\%  & 4.37 \\ 
    	      &      &      & [ArIII]   & 7135  & 60.7 & 20\%  & 12.3 \\
    	      &      &      & [ArIII]   & 7751  & 18.6 & 26\%  & 2.64 \\
\hline
HL90-50 & 880.& 1.52& [OII]     & 3727  & 43.0  & 08\% &  115. \\ 
        &     &     & [NeIII]	& 3869  & 8.31  & 11\% &  19.6 \\
        &     &     & HeI	& 3889  & 8.28  & 11\% &  19.2 \\
        &     &     & [NeIII]/HI& 3968  & 9.47  & 11\% &  20.5 \\
        &     &     & HI	& 4100  & 18.8  & 09\% &  36.5 \\
        &     &     & HI	& 4340  & 39.7  & 08\% &  62.6 \\
        &     &     & [OIII]	& 4363  & 2.71  & 10\% &  4.20 \\
        &     &     & HeI	& 4471  & 4.22  & 13\% &  5.94 \\
        &     &     & HI	& 4861  & 100.  & 07\% &  100. \\
        &     &     & [OIII]	& 4959  & 130.  & 06\% &  120. \\
        &     &     & [OIII]	& 5007  & 390.  & 04\% &  344. \\
        &     &     & HeI	& 5876  & 33.5  & 08\% &  16.3 \\
        &     &     & [OI]	& 6300  & 1.06  & 14\% &  0.42 \\
        &     &     & [NII]	& 6548  & 17.4  & 04\% &  6.09 \\
        &     &     & HI	& 6563  & 823.  & 08\% &  285. \\
        &     &     & [NII]	& 6584  & 48.5  & 08\% &  16.6 \\
        &     &     & [SII]	& 6717  & 31.5  & 09\% &  10.1 \\
        &     &     & [SII]	& 6731  & 24.9  & 09\% &  7.96 \\
        &     &     & HeI	& 7065  & 8.56  & 08\% &  2.32 \\
        &     &     & [ArIII]	& 7135  & 48.1  & 11\% &  12.6 \\
        &     &     & [OII]	& 7320  & 6.88  & 11\% &  1.65 \\
        &     &     & [OII]	& 7330  & 6.39  & 09\% &  1.52 \\
        &     &     & [ArIII]	& 7751  & 12.7  & 09\% &  2.48 \\
        &     &     & [SIII]	& 9069  & 49.0  & 08\% &  5.03 \\
\hline
HL90-51& 47. & 1.88& HI        	& 4100  & 20.4  & 16\% &  46.4 \\
       &     &     & HI        	& 4340  & 25.5  & 15\% &  44.9 \\
       &     &     & [OIII]    	& 4363  & 2:  & -    &   4.1  \\   
       &     &     & HI        	& 4861  & 100.  & 10\% &  100. \\
       &     &     & [OIII]    	& 4959  & 151.  & 09\% &  137. \\    
       &     &     & [OIII]    	& 5007  & 397.  & 05\% &  340. \\  
       &     &     & HeI       	& 5876  & 26.1  & 15\% &  10.7 \\
       &     &     & [OI]      	& 6300  & 2.6   & 20\% &  1.   \\
       &     &     & [NII]     	& 6548  & 22.7  & 16\% &  6.2  \\
       &     &     & HI        	& 6563  & 1063. & 07\% &  285. \\
       &     &     & [NII]     	& 6584  & 71.3  & 11\% &  19.  \\ 
       &     &     & [SII]     	& 6717  & 51.2  & 12\% &  12.6 \\  
       &     &     & [SII]     	& 6731  & 47.8  & 12\% &  11.7 \\  
       &     &     & HeI       	& 7065  & 21.6  & 16\% &  4.3  \\  
       &     &     & [ArIII]   	& 7135  & 63.1  & 11\% &   12. \\
       &     &     & [OII]     	& 7320  & 16.2  & 16\% &   2.8 \\
       &     &     & [OII]     	& 7330  & 9.81  & 17\% &   1.7 \\ 
       &     &     & [ArIII]   	& 7751  & 17.7  & 16\% &   2.3 \\
       &     &     & [SIII]    	& 9069  & 115.  & 10\% &   8.9 \\ 

\hline
\end{tabular}
}
\end{minipage}
\label{tabHII_flux}
\end{table}
\begin{table}
\centering

\begin{minipage}{85mm}
{\scriptsize  
\contcaption{}
\begin{tabular}{@{}ccclcrrr@{}}
\hline
ID & F$_{{\rm H}\beta}$ & c({H$\beta$}) & Ion & $\lambda$ (\AA) & F$_{\lambda}$ & $\Delta$F$_{\lambda}$ & I$_{\lambda}$ \\ 
\hline
HL90-107& 57. & 1.74& HI	& 4340  & 21.2 & 8.5\% & 35.7 \\
        &     &     & [OIII]	& 4363  & 5.6: &  -    & 9.3: \\
        &     &     & HI	& 4861  & 100. & 06\%  & 100. \\
        &     &     & [OIII]	& 4959  & 285. & 05\%  & 261. \\
        &     &     & [OIII]	& 5007  & 862. & 03\%  & 751. \\
        &     &     & HeI	& 5876  & 26.9 & 08\%  & 11.7 \\
        &     &     & [OI]	& 6300  & 12.7 & 8.5\% & 4.39 \\
        &     &     & [OI]	& 6363  & 4.95 & 10\%  & 1.64 \\
        &     &     & [NII]	& 6548  & 22.4 & 08\%  & 6.70 \\
        &     &     & HI	& 6563  & 963. & 03\%  & 285. \\
        &     &     & [NII]	& 6584  & 70.2 & 06\%  & 20.5 \\ 
        &     &     & HeI	& 6678  & 13.8 & 8.5\% & 3.90 \\
        &     &     & [SII]	& 6717  & 100. & 06\%  & 27.8 \\
        &     &     & [SII]	& 6731  & 79.3 & 06\%  & 21.7 \\ 
        &     &     & HeI	& 7065  & 14.4 & 8.5\% & 2.49 \\
        &     &     & [ArIII]	& 7135  & 75.8 & 06\%  & 16.3 \\
        &     &     & [OII]	& 7320  & 8.30 & 8.5\% & 1.60 \\ 
        &     &     & [OII]	& 7330  & 8.19 & 8.5\% & 1.60 \\ 
        &     &     & [ArIII]	& 7751  & 18.4 & 08\%  & 2.82 \\
        &     &     & [SIII]	& 9069  & 98.9 & 06\%  & 7.33 \\ 
\hline
HL90-111& 4300.& 1.05& [OII]    & 3727  & 71.6   & 08\%  &   141. \\ 
    	&      &     & HI	& 4340  & 26.5   & 10\%  &   36.3 \\
    	&      &     & [OIII]	& 4363  & 0.93:  & -	 &   1.26:\\
    	&      &     & HI	& 4861  & 100.   & 06\%  &   100. \\
    	&      &     & [OIII]	& 4959  & 65.7   &08\%   &   62.2 \\
    	&      &     & [OIII]	& 5007  & 196.5  & 4.5\% &   180. \\
    	&      &     & HeI	& 5876  & 18.0   &10\%   &   10.9 \\
    	&      &     & [NII]	& 6548  & 19.3   &10\%   &   9.33 \\
    	&      &     & HI	& 6563  & 593.   &04\%   &   285. \\
    	&      &     & [NII]	& 6584  & 57.0   &08\%   &   27.1 \\
    	&      &     & HeI	& 6678  & 8.56   & 13\%  &   3.96 \\
    	&      &     & [SII]	& 6717  & 37.1   &8.5\%  &   16.9 \\
    	&      &     & [SII]	& 6731  & 28.0   &10\%   &   12.7 \\
    	&      &     & HeI	& 7065  & 5.97   & 13\%  &   2.42 \\
    	&      &     & [ArIII]	& 7135  & 31.4   &13\%   &   12.4 \\
    	&      &     & [OII]	& 7320  & 7.15   & 13\%  &   2.66 \\
    	&      &     & [OII]	& 7330  & 5.92   & 13\%  &   2.19 \\
    	&      &     & [ArIII]	& 7751  & 8.43   & 13\%  &   2.72 \\
\hline
HL90-120& 108.& 0.40& [OII]     & 3727  & 41.6   & 11\% &  48.4  \\ 
    	&     &     & HI	& 3835  & 3.22   & 16\% &  3.70  \\
    	&     &     & [NeIII]	& 3869  & 14.9   & 13\% &  17.0  \\	
    	&     &     & HeI	& 3889  & 9.45   & 14\% &  10.7  \\   
    	&     &     & [NeIII]/HI& 3968  & 13.9   & 13\% &  15.7  \\   
    	&     &     & HI	& 4100  & 16.1   & 13\% &  17.9  \\
    	&     &     & HI	& 4340  & 36.1   & 11\% &  38.7  \\
    	&     &     & [OIII]	& 4363  & 1.69   & 16\% &  1.81  \\   
    	&     &     & HeI	& 4471  & 3.60   & 16\% &  3.80  \\ 
    	&     &     & HeII	& 4686  & 4.82   & 16\% &  4.94 \footnote{Stellar broad line emission from WR stars}  \\
    	&     &     & HI	& 4861  & 100.   & 10\%  &  100. \\
    	&     &     & [OIII]	& 4959  & 149.   & 10\%  &  147. \\	
    	&     &     & [OIII]	& 5007  & 378.   & 09\%  &  370. \\  
    	&     &     & HeI	& 5876  & 5.89   & 16\%  &  5.27 \\
    	&     &     & [OI]	& 6300  & 1.27   & 16\%  &  1.10 \\
    	&     &     & [SIII]	& 6312  & 1.54   & 16\%  &  1.2  \\
    	&     &     & [NII]	& 6548  & 6.95   & 15\%  &  5.3  \\
    	&     &     & HI	& 6563  & 376.   & 09\%  &  285. \\
    	&     &     & [NII]	& 6584  & 23.0   & 12.5\%& 17.4  \\  
    	&     &     & HeI	& 6678  & 2.31   & 16\%  &  1.7  \\
    	&     &     & [SII]	& 6717  & 12.9   & 13\%  &  9.7  \\
    	&     &     & [SII]	& 6731  & 9.55   & 14\%  &  7.1  \\
    	&     &     & HeI	& 7065  & 1.51   & 16\%  &  1.1  \\
    	&     &     & [ArIII]	& 7135  & 9.54   & 14\%  &  6.7  \\
    	&     &     & [OII]	& 7320  & 2.23   & 16\%  &  1.5  \\
    	&     &     & [OII]	& 7330  & 1.73   & 16\%  &  1.2  \\
\hline
\end{tabular}
}
\end{minipage}
\label{tabHII_flux}
\end{table}

\begin{table}
\centering
\begin{minipage}{85mm}
{\scriptsize  
\contcaption{}
\begin{tabular}{@{}ccclcrrr@{}}
\hline
ID & F$_{{\rm H}\beta}$ & c({H$\beta$}) & Ion & $\lambda$ (\AA) & F$_{\lambda}$ & $\Delta$F$_{\lambda}$ & I$_{\lambda}$ \\ 
\hline
HL90-122& 61. & 1.04& HI        & 4340  & 29.7 & 10\%  & 40.6 \\
    	&     &     & HI	& 4861  & 100. & 09\%  & 100. \\
    	&     &     & [OIII]	& 4959  & 21.6 & 11\%  & 20.4 \\
    	&     &     & [OIII]	& 5007  & 64.7 & 9.5\% & 59.5 \\
    	&     &     & HeI	& 5876  & 5.97 & 16\%  & 3.64 \\
    	&     &     & [NII]	& 6548  & 30.2 & 10\%  & 14.7 \\
    	&     &     & HI	& 6563  & 591. & 06\%  & 286. \\
    	&     &     & [NII]	& 6584  & 93.5 & 09\%  & 44.9 \\
    	&     &     & HeI	& 6678  & 4.48 & 17\%  & 2.08 \\
    	&     &     & [SII]	& 6717  & 61.9 & 9.5\% & 28.4 \\
    	&     &     & [SII]	& 6731  & 45.6 & 10\%  & 20.8 \\
    	&     &     & HeI	& 7065  & 3.24 & 18\%  & 1.33 \\
    	&     &     & [ArIII]	& 7135  & 16.1 & 11\%  & 6.44 \\
    	&     &     & [OII]	& 7320  & 5.16 & 16\%  & 1.94 \\
    	&     &     & [OII]	& 7330  & 5.64 & 16\%  & 2.11 \\
\hline
HL90-127& 12.4& 1.10& HI        & 4340  & 26.5  & 10\% & 36.9 \\
    	&     &     & HI	& 4861  & 100.  & 07\% & 100. \\
    	&     &     & [OIII]	& 4959  & 158.  & 06\% & 150. \\
    	&     &     & [OIII]	& 5007  & 477.  &05\%  & 437. \\
    	&     &     & HeI	& 5876  & 21.3  &10\%  & 12.6 \\
    	&     &     & [NII]	& 6548  & 9.49  &12\%  & 4.43 \\
    	&     &     & HI	& 6563  & 616.  &04\%  & 285. \\
    	&     &     & [NII]	& 6584  & 29.2  &10\%  & 13.4 \\
    	&     &     & HeI	& 6678  & 12.1  &12\%  & 5.42 \\
    	&     &     & [ArIII]	& 7135  & 35.3  &09\%  & 13.3 \\
\hline
HL90-141& 48.7& 1.23& HI        & 4340  & 26.7  & 15\% &  38.7 \\
    	&     &     & HI	& 4861  & 100.  & 08\% &  100. \\
    	&     &     & [OIII]	& 4959  & 66.7  & 08\% &  62.6 \\
    	&     &     & [OIII]	& 5007  & 210.  & 05\% &  190. \\ 
    	&     &     & [NII]	& 6548  & 22.0  & 15\% &  9.38 \\
    	&     &     & HI	& 6563  & 675.  & 05\% &  286. \\
    	&     &     & [NII]	& 6584  & 87.3  & 08\% &  36.6 \\
    	&     &     & HeI	& 6678  & 9.00  & 18\% &  3.64 \\
    	&     &     & [SII]	& 6717  & 64.2  & 08\% &  25.5 \\ 
    	&     &     & [SII]	& 6731  & 44.5  & 09\% &  17.6 \\ 
    	&     &     & [ArIII]	& 7135  & 24.0  & 15\% &  8.12 \\
    	&     &     & [OII]	& 7320  & 5.64  & 25\% &  1.77 \\
    	&     &     & [OII]	& 7330  & 9.62  & 18\% &  3.00 \\
\hline
\end{tabular}
}
\end{minipage}
\label{tabHII_flux}
\end{table}

\end{document}